\documentclass[twoside]{article}
\makeatletter\if@twocolumn\PassOptionsToPackage{switch}{lineno}\else\fi\makeatother

\usepackage{graphicx}
\usepackage{caption}
\usepackage{amsmath}
\usepackage{fancyhdr}
\usepackage{latexsym}
\usepackage{tabulary}
\usepackage[utf8]{inputenc}
\usepackage{url,multirow,morefloats,floatflt,cancel,tfrupee}
\makeatletter

\AtBeginDocument{\@ifpackageloaded{textcomp}{}{\usepackage{textcomp}}}
\makeatother
\usepackage{colortbl}
\usepackage{xcolor}
\usepackage{pifont}
\usepackage[nointegrals]{wasysym}
\makeatletter

\def\mcWidth#1{\csname TY@F#1\endcsname+\tabcolsep}

\def\cAlignHack{\rightskip\@flushglue\leftskip\@flushglue\parindent\z@\parfillskip\z@skip}
\def\rAlignHack{\rightskip\z@skip\leftskip\@flushglue \parindent\z@\parfillskip\z@skip}

\@ifundefined{etal}{}{}

\usepackage{ifxetex}
\ifxetex\else\if@twocolumn\@ifpackageloaded{stfloats}{}{\usepackage{dblfloatfix}}\fi\fi

\AtBeginDocument{
\expandafter\ifx\csname eqalign\endcsname\relax
\def\eqalign#1{\null\vcenter{\def\\{\cr}\openup\jot\m@th
  \ialign{\strut$\displaystyle{##}$\hfil&$\displaystyle{{}##}$\hfil
      \crcr#1\crcr}}\,}
\fi
}

\AtBeginDocument{%
  \@ifpackageloaded{endfloat}%
   {\renewcommand\efloat@iwrite[1]{\immediate\expandafter\protected@write\csname efloat@post#1\endcsname{}}}{\newif\ifefloat@tables}%
}%

\def\BreakURLText#1{\@tfor\brk@tempa:=#1\do{\brk@tempa\hskip0pt}}
\let\lt=<
\let\gt=>
\def\processVert{\ifmmode|\else\textbar\fi}

\@ifundefined{subparagraph}{
\def\subparagraph{\@startsection{paragraph}{5}{2\parindent}{0ex plus 0.1ex minus 0.1ex}%
{0ex}{\normalfont\small\itshape}}%
}{}

\newcommand\role[1]{\unskip}
\newcommand\aucollab[1]{\unskip}
  
\@ifundefined{tsGraphicsScaleX}{\gdef\tsGraphicsScaleX{1}}{}
\@ifundefined{tsGraphicsScaleY}{\gdef\tsGraphicsScaleY{.9}}{}
\def\checkGraphicsWidth{\ifdim\Gin@nat@width>\linewidth
	\tsGraphicsScaleX\linewidth\else\Gin@nat@width\fi}

\def\checkGraphicsHeight{\ifdim\Gin@nat@height>.9\textheight
	\tsGraphicsScaleY\textheight\else\Gin@nat@height\fi}

\def\fixFloatSize#1{}
\let\ts@includegraphics\includegraphics

\def\inlinegraphic[#1]#2{{\edef\@tempa{#1}\edef\baseline@shift{\ifx\@tempa\@empty0\else#1\fi}\edef\tempZ{\the\numexpr(\numexpr(\baseline@shift*\f@size/100))}\protect\raisebox{\tempZ pt}{\ts@includegraphics{#2}}}}

\AtBeginDocument{\def\includegraphics{\@ifnextchar[{\ts@includegraphics}{\ts@includegraphics[width=\checkGraphicsWidth,height=\checkGraphicsHeight,keepaspectratio]}}}

\DeclareMathAlphabet{\mathpzc}{OT1}{pzc}{m}{it}

\def\URL#1#2{\@ifundefined{href}{#2}{\href{#1}{#2}}}

\def\UrlOrds{\do\*\do\-\do\~\do\'\do\"\do\-}%
\g@addto@macro{\UrlBreaks}{\UrlOrds}

\edef\fntEncoding{\f@encoding}

\makeatother

\newif\ifmultipleabstract\multipleabstractfalse%
\usepackage[paperheight=11in,paperwidth=8.3in,margin=2.5cm,headsep=.7cm,top=2.5cm]{geometry}
\usepackage[T1]{fontenc}

\renewenvironment{abstract}
{\vspace*{-1pc}\trivlist\item[]\leftskip\hindawiIndent\par\vskip4pt\noindent\textbf{\abstractname}\mbox{\null}\\}{\par\noindent\endtrivlist}

 \date{} \emergencystretch 8pt

\makeatletter\def\hindawiIndent{0pc}
\def\author#1{\gdef\@author{\hskip-\dimexpr(\tabcolsep)\hskip\hindawiIndent\parbox{\dimexpr\textwidth-\hindawiIndent}{\raggedright\bfseries#1}}}

\def\title#1{\gdef\@title{\vspace*{-30pt}\raggedright\textbf{ \journaltitle}~\\\raggedright\bfseries\ifx\@articleType\@empty\vspace*{20pt}\else\vspace*{20pt}\@articleType\vspace*{20pt}\\\fi#1}}
\let\@articleType\@empty \def\articletype#1{\gdef\@articleType{{\normalfont\itshape#1}}}

\let\@runningHead\@empty \def\RunningHead#1{\gdef\@runningHead{{\normalfont #1}}}

\usepackage{fancyhdr}
\makeatother

\newlength\savedwidth

\newcommand\thickhline{\noalign{\global\savedwidth\arrayrulewidth\global\arrayrulewidth 2pt}%
\hline
\noalign{\global\arrayrulewidth\savedwidth}}

\usepackage[numbers,sort&compress]{natbib}
\usepackage{enumitem}
\usepackage{soul}
\usepackage{makecell}
\usepackage{multirow}
\usepackage{subfig}
\usepackage{xcolor}
\usepackage[colorlinks = true, linkcolor = blue, urlcolor  = blue, citecolor = blue, anchorcolor = blue]{hyperref}

\def\journaltitle{}
\newcommand\blfootnote[1]{%
  \begingroup
  \renewcommand\thefootnote{}\footnote{#1}%
  \addtocounter{footnote}{-1}%
  \endgroup
}

\catcode`\_=11\relax
    \newcommand\email[1]{\_email #1\q_nil}
    \def\_email#1@#2\q_nil{%
      \href{mailto:#1@#2}{{\color{black}\emailfont #1\emailampersat #2}}
    }
    \newcommand\emailfont{\ttfamily}
    \newcommand\emailampersat{{\color{black}\small@}}
    \catcode`\_=8\relax

\begin{document}

\nocite{*}

\title{Evaluation of Machine Learning Algorithms in Network-Based Intrusion Detection System}
\author{Tuan-Hong Chua 
		and Iftekhar Salam \blfootnote{Email addresses: \email{swe1809792@xmu.edu.my} (Tuan-Hong Chua), \email{iftekhar.salam@xmu.edu.my} (Iftekhar Salam)}~\\[-3pt]\normalsize\normalfont \mbox{}\\{School of Computing and Data Science\unskip, Xiamen University Malaysia\unskip, Sepang 43900\unskip, Malaysia}
		}

\maketitle
\begin{abstract}
Cybersecurity has become one of the focuses of organisations. The number of cyberattacks keeps increasing as Internet usage continues to grow. An intrusion detection system (IDS) is an alarm system that helps to detect cyberattacks. As new types of cyberattacks continue to emerge, researchers focus on developing machine learning (ML)-based IDS to detect zero-day attacks. Researchers usually remove some or all attack samples from the training dataset and only include them in the testing dataset when evaluating the performance of an IDS on detecting zero-day attacks. Although this method may show the ability of an IDs to detect unknown attacks; however, it does not reflect the long-term performance of the IDS as it only shows the changes in the type of attacks. In this paper, we focus on evaluating the long-term performance of ML-based IDS. To achieve this goal, we propose evaluating the ML-based IDS using a dataset that is created later than the training dataset. The proposed method can better assess the long-term performance of an ML-based IDS, as the testing dataset reflects the changes in the type of attack and the changes in network infrastructure over time. We have implemented six of the most popular ML models that are used for IDS, including decision tree (DT), random forest (RF), support vector machine (SVM), naïve Bayes (NB), artificial neural network (ANN) and deep neural network (DNN). Our experiments using the CIC-IDS2017 and the CSE-CIC-IDS2018 datasets show that SVM and ANN are most resistant to overfitting. Besides that, our experiment results also show that DT and RF suffer the most from overfitting, although they perform well on the training dataset. On the other hand, our experiments using the LUFlow dataset have shown that all models can perform well when the difference between the training and testing datasets is small. 
\end{abstract}

\section{Introduction}
Cybersecurity has gained more attention in recent years as we rely more on computers. Due to the global pandemic, our lives are moving online; however, at the same time, cybersecurity issues are getting even worse. We rely more on the Internet for daily activities, including virtual meetings, purchasing daily necessities, and ordering food. At the same time, cyberattacks are also skyrocketing. According to a report from PR Newswire, the FBI reported that the cyberattacks on their Cyber Division had increased by 400\%, to nearly 4000 attacks per day \citep{monstercloud2020top} since the global pandemic. Organisations should take necessary countermeasures to address the cyberattacks. There are multiple countermeasures available to safeguard the security of an organisation’s computer system, including network access control, antivirus software and Virtual Private Network (VPN). Besides that, an Intrusion Detection System (IDS) is also one of the control measures. 
An intrusion detection system (IDS) is a hardware or software application that ensures the security of computer systems by monitoring traffic for the sign of intrusions \citep{liao2013intrusion}. Once suspicious activities have been identified, an alarm will be raised, and IT personnel could take action accordingly. Depending on the monitored traffic, an IDS can be classified as network-based IDS (NIDS) or host-based IDS (HIDS) \citep{liao2013intrusion}. A NIDS monitors the network traffic while HIDS monitors operating system files. In literature, NIDS is usually referred to as flow-based NIDS, where only the packet header is analysed to detect intrusions.

Besides NIDS and HIDS, an IDS can also be classified as signature-based IDS or anomaly-based IDS \citep{liao2013intrusion}, based on the method that is used to detect intrusions. A signature-based IDS is also known as misuse detection \citep{khraisat2018anomaly}, where attacks are identified by utilising the signature of known attacks stored in the database. Hence, it cannot capture attacks that it has never seen before. The advantage of signature-based IDS is zero false-positive rates, as it will never classify a benign activity as malicious \citep{kreibich2004honeycomb}. On the other hand, anomaly-based IDS classify network traffic with a set of pre-defined rules for “normal activity”. If an activity does not fit into those rules, it will be classified as “suspicious activity” \citep{garcia2009anomaly}. Therefore, an anomaly-based IDS is also known as rule-based IDS \citep{liao2013intrusion}. The main distinction from signature-based IDS is that it identifies attacks based on traffic behaviour instead of explicit signatures. This gives the anomaly-based IDS the flexibility to identify unseen attacks. 

In recent years, existing works focused heavily on adapting Machine Learning (ML) to improve the accuracy of IDS. ML is a science that aims to imitate human’s ability to learn \citep{geron2019hands}. When used for IDS, it can learn the behaviour of benign and malicious network traffic and differentiate between them. Recent research has proven that IDS that adopts ML algorithms can achieve good accuracy and surpass conventional methods. One weakness that the IDS typically suffers from is detecting unseen attacks, specifically zero-day attacks. A zero-day attack occurs when the hackers exploit a system vulnerability that is unknown to the developer or before the developers address it \citep{kaspersky2021}. Although having high accuracy in detecting known attack activities, IDS often has the limitation of detecting zero-day attacks and unseen attacks. The weakness might have been solved by utilising Machine Learning (ML). Research done by Hindy et al. \citep{hindy2020utilising} has shown promising results in detecting zero-day attacks by using Support Vector Machine (SVM) and Artifical Neural Network (ANN) for IDS. 

\begin{figure}[!htbp]
\centering
\includegraphics[width=0.8\textwidth]{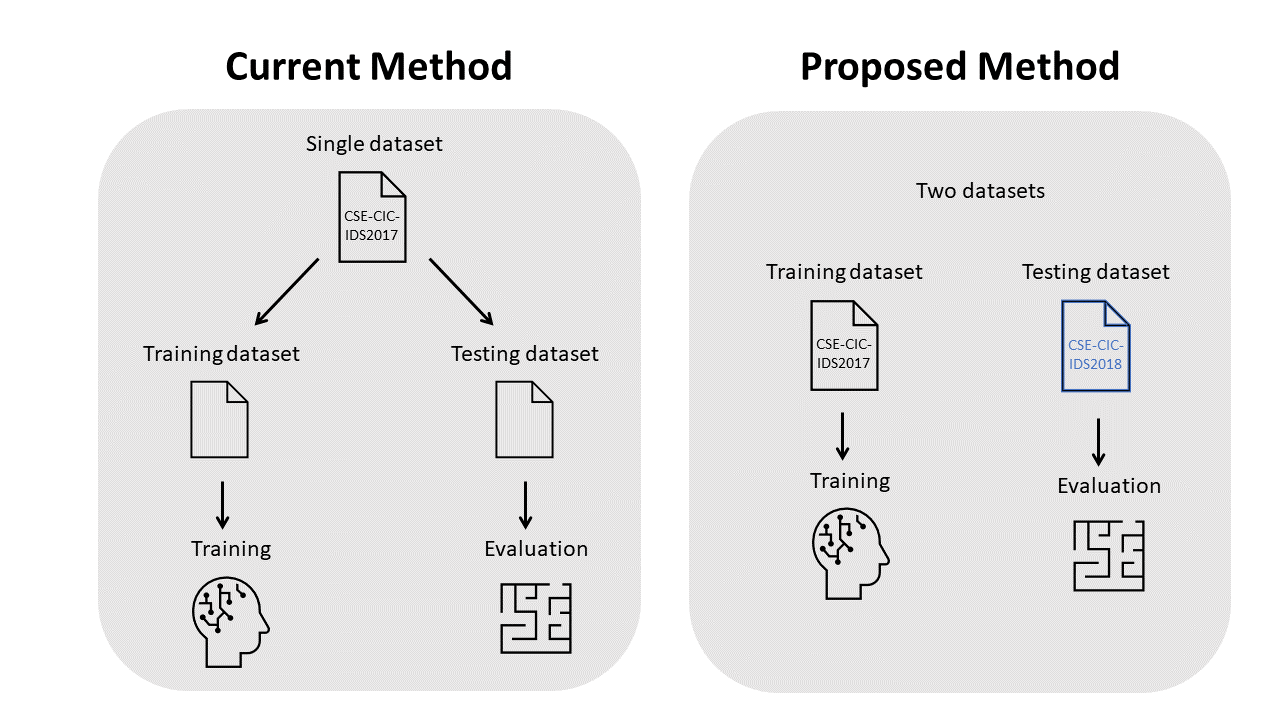}
\caption{Current studies used the same dataset for both training and testing while this paper proposed to use different datasets for training and testing.}
\label{fig:1}
\end{figure}

Although studies in recent years have achieved compelling results, one particular gap in those studies is that a single dataset is being used for both training and evaluation. For example, the study by Hindy et al. \citep{hindy2020utilising} used only the normal traffic for training, while the attack activities were used to mimic zero-day attacks. However, we are not able to get a clear picture of the long-term performance of an IDS when the same dataset is being used for both training and testing. In a real-world scenario, we usually have to train that model using an existing dataset while using the model to examine future network traffic. The point is that the network environment in the future will not be the same as today. For instance, organisations and attackers will update their infrastructures and network topologies over time. Besides that, zero-day attacks in the future will never be available in the existing dataset. The interest of this paper is to further increase the deviation between the data used for training and the data used for evaluation to mimic real-world situations. Thus, we use different datasets to train and evaluate our ML models. The difference between the existing methods and the proposed method is further illustrated in Figure~\ref{fig:1}.
Based on the proposed method, our contributions in this paper are listed as below:
\begin{enumerate}[label=(\alph*),noitemsep]
\item We proposed training and evaluating ML-based IDS using different datasets to mimic real-world scenarios. 
\item We identified multiple datasets for the evaluation, including the CIC-IDS2017, the CSE-CIC-IDS2018, and the LUFlow datasets. 
\item We compared the performance of decision tree (DT), random forest (RF), support vector machine (SVM), naïve Bayes (NB), artificial neural network (ANN) and deep neural network (DNN) using the proposed method. 
\end{enumerate}
The remainder of the paper is organised as follows: Section~\ref{sec:related_background} provides some related background. In Section~\ref{sec:related_works}, we review related literature. The experiment process and experiment environment are described in Section~\ref{sec:framework_of_experiment}. Section~\ref{sec:experiment_and_discussion} discusses the experiment results, and further discussions on the results are provided in Section~\ref{sec:discussion_of_results}. Finally, Section~\ref{sec:conclusion} concludes this paper.

\section{Related Background}
\label{sec:related_background}
In this section, we introduce some of the most well-known datasets that are used in the domain of IDS. We also discuss the machine learning models that are used in this paper.
\subsection{Dataset}
\label{sec:dataset}
The availability of datasets is one of the biggest challenges in the domain of IDS. Due to privacy and security reasons, most organisations will never share their network traffic data. However, a high-quality dataset is crucial to develop an anomaly-based IDS and evaluate its performance. Hence, multiple datasets have been developed by different organisations for research purposes. In this section, we introduce the datasets that are used for our experiments.
\subsubsection{CIC-IDS2017 Dataset}
The CIC-IDS2017 dataset \citep{sharafaldin2018toward} is one of the most popular datasets from recent years. The dataset was created by the Canadian Institute for Cybersecurity (CIC) in 2017. As the dataset is created recently, it covers various operating systems, protocols and attack types. A comprehensive network environment consisting of modem, firewall, switches and routers, with different operating systems, including Windows, Mac OS and Ubuntu, was set up to create the dataset. The behaviour of 25 users was simulated, and protocols like HTTP, HTTPS, FTP, SSH and email protocols were used to develop benign traffic. After that, the most common attacks in 2016 were simulated. 
The attacks include Brute force attack, DoS attack, DDoS attack, Infiltration attack, Heart-bleed attack, Botnet attack and Port Scan attack. The dataset was then made publicly available on the University of New Brunswick’s website \citep{web:cic2017} as a CSV file. The biggest downside of the dataset is that it suffers from the high-class imbalance problem \citep{thakkar2020review}, where more than 70\% of the traffics are benign, and some of the attacks contribute to less than 1\% of the overall traffic.
\subsubsection{CSE-CIC-IDS2018 Dataset}
The CSE-CIC-IDS2018 is an updated version of the CIC-IDS2017 dataset. The dataset is created by CIC in conjunction with the Communications Security Establishments (CSE). The dataset is publicly accessible~\citep{web:cic2018} through Amazon Web Service (AWS) and is widely used in recent studies. The most significant update is the laboratory environment that is used to create the dataset; it is significantly more robust than its predecessor. Fifty machines were used to attack the infrastructure, and the victim organisation was simulated with 420 machines and 30 servers across five departments. Besides that, different versions of the Windows operating system were installed on the victim’s devices, including Windows 8.1 and Windows 10. The servers were also running on different server operating systems, including Windows server 2012 and Windows server 2016 \citep{web:cic2018aws}. The diversity of the devices in the laboratory environment makes it more similar to real-world networks. The attacks towards the infrastructure are similar to the CIC-IDS2017 dataset. 
\subsubsection{LUFlow Dataset}
The LUFlow dataset \citep{mills2021practical} is one of the latest datasets for the domain of IDS. Lancaster University created the dataset in 2020, and it was still being updated in 2021. The LUFlow dataset is unique because the data is real-world data, which is captured through the honeypots within Lancaster University’s public address space. Hence, the dataset can reflect the emerging threats in the real world. The limitation of the dataset is that it only labels the traffic as benign or malicious instead of an explicit type of attack. This limitation is due to the fact that the data is collected from the real world instead of a laboratory environment; hence, it is impossible to tell the real intention behind each traffic. Besides that, some traffics are classified as “outlier” if the traffics contain suspicious activity but without connection to a malicious node. The existence of the outlier label indicates that unknown attacks may be classified as outliers instead of malicious and might cause the dataset to be less meaningful.
\subsection{Machine Learning Models}
Machine learning (ML) is getting more and more attention in recent years, and researchers are exploring its application in different areas. ML models are great at predicting or classifying data \citep{geron2019hands}. In the context of IDS, ML is adopted to classify whether the traffic is benign or an attack. In this section, we briefly discuss some of the most widely used ML models in the domain of IDS. 
\subsubsection{Decision Tree}
\begin{figure}[!htbp]
\centering
\includegraphics[width=0.75\textwidth]{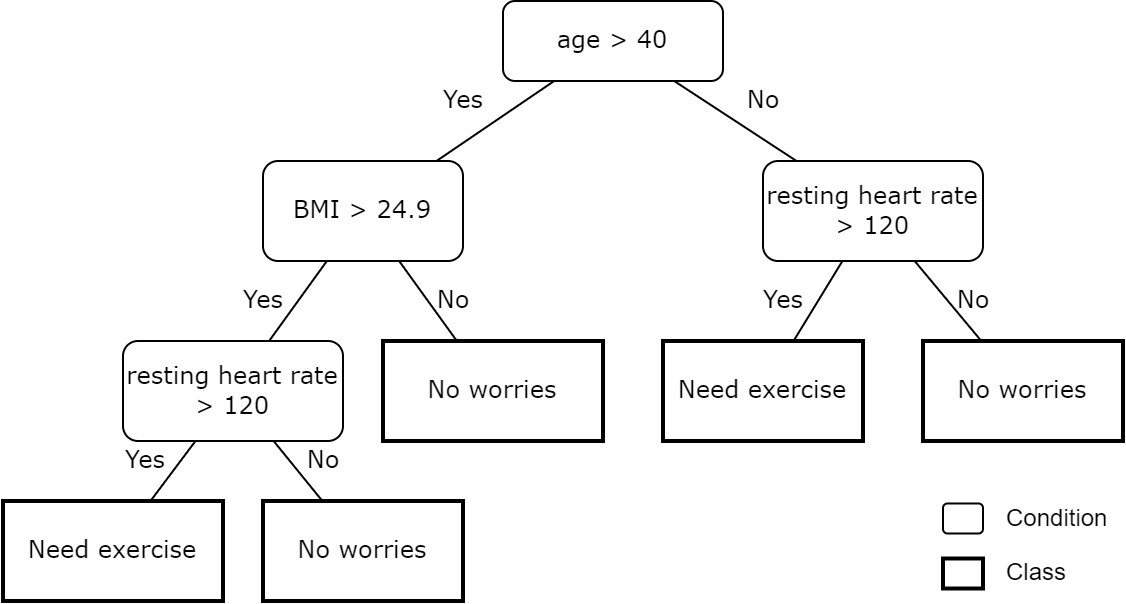}
\caption{An example of a Decision Tree.}
\label{fig:2}
\end{figure}

A decision tree (DT) classifier is a model that classifies the sample using a tree-like structure. Each decision tree is made up of multiple nodes and leaves, where the nodes represent some conditions while the leaves represent different classes \citep{safavian1991survey}. A decision tree will be built during the training stage based on the training data. Figure~\ref{fig:2} shows an example of the creation process of a decision tree. This is an iterative process. 
In each iteration, one feature will be selected to split the data on a leaf node until all data in a leaf node are homogeneous. The selection of features is based on how well a feature can split the data into a homogeneous group, measured using metrics like Gini and Entropy. The features closer to the root node split the data better and have a higher correlation with the predicted output. Therefore, the decision tree can facilitate feature selection. In the testing stage, each sample in the testing dataset will be input to the tree and traverse from the root node, i.e., the node at the top of the tree, to a leaf node. The leaf node to which a sample end represents the class of that sample. The decision tree is one of the most popular models in the domain of IDS as it provides good accuracy while being easy to train. 

\subsubsection{Random Forest}
\begin{figure}[!htbp]
\centering
\includegraphics[width=0.75\textwidth]{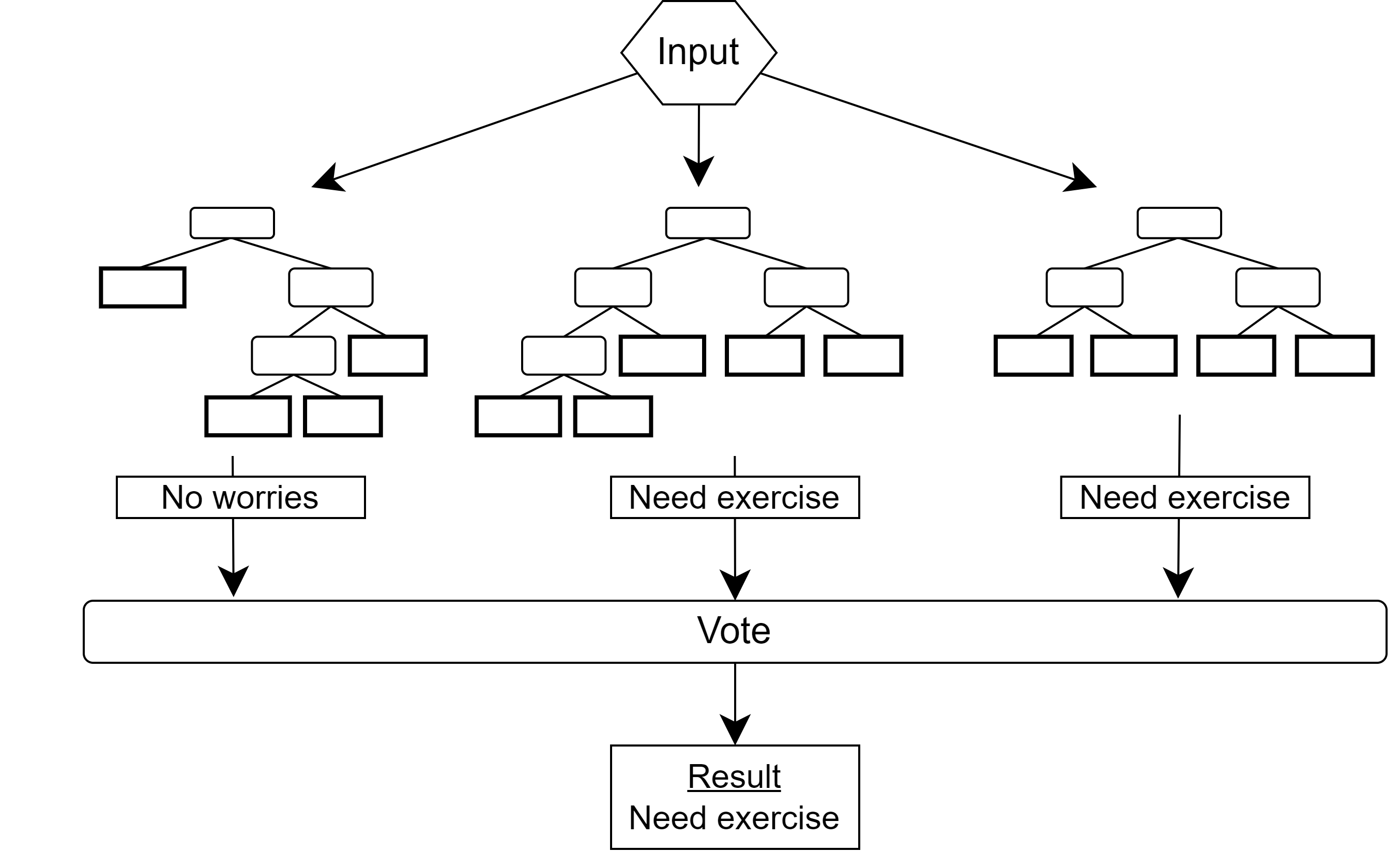}
\caption{Illustration of a simple Random Forest with three trees.}
\label{fig:3}
\end{figure}

Random forest (RF) \citep{breiman2001random} is a model made up of a branch of decision trees. A bootstrapped dataset is created to build each decision tree within the random forest by randomly picking rows from the original training dataset. Since each decision tree is trained using a different bootstrapped dataset, each is slightly different from the other. As illustrated in Figure~\ref{fig:3}, the votes from each tree are collected, and the classification depends on the class that received the most votes. By involving multiple decision trees, random forest is less prone to overfitting problems and less sensitive to the noise in the dataset while maintaining the simplicity of the decision tree \citep{breiman2001random}.
\subsubsection{Support Vector Machine}
\begin{figure}[!htbp]
\centering
\includegraphics[width=0.3\textwidth]{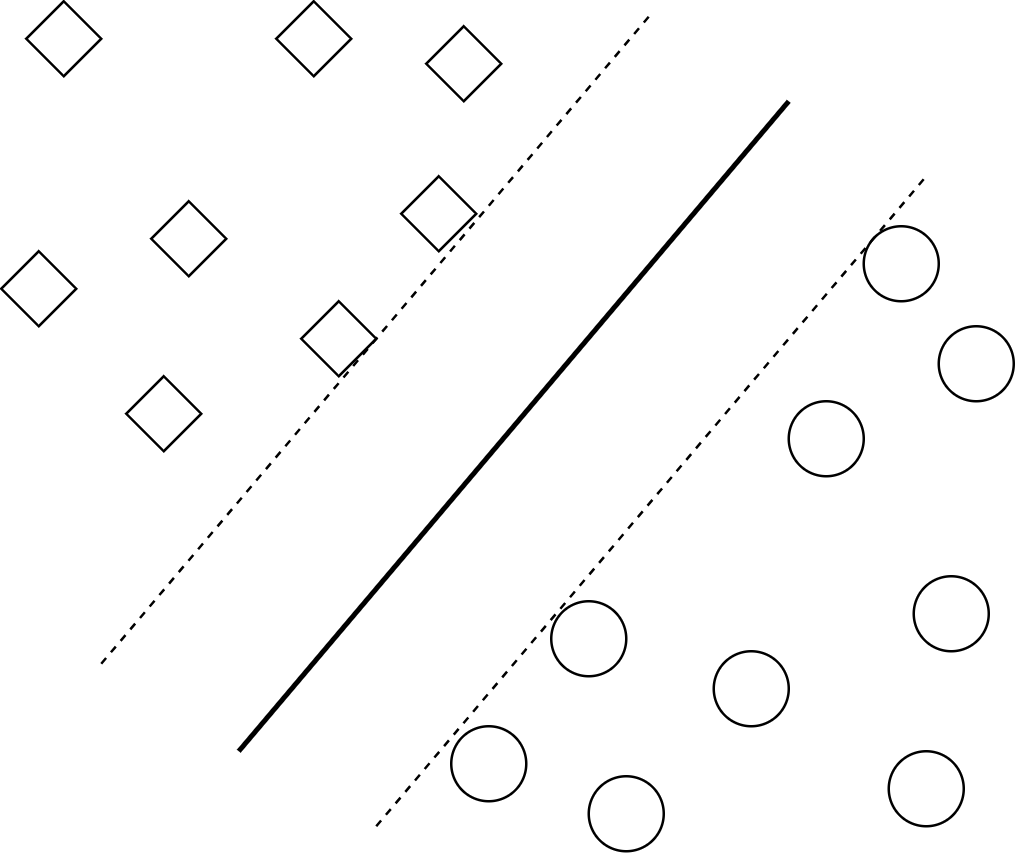}
\caption{A simple SVM in a 2-dimensional space.}
\label{fig:4}
\end{figure}

Support Vector Machine (SVM) is a model that aims to define a hyperplane that can separate the training data into its respective classes \citep{cortes1995support}. As illustrated in Figure~\ref{fig:4}, the thick line represents a separator in a 2-dimensional space that separates the data of two different classes. To prevent overfitting, the separator in Figure~\ref{fig:4} should be in the middle of the dashed lines to ensure that the distance to data points of either class is maximum. %
One challenge of SVM is that the data might be inseparable in the input space. To solve this issue, kernel functions are used to non-linearly map the data to a higher dimensional feature space \citep{cortes1995support}. If the data is still inseparable in the higher dimensional space, the model will map the data to even higher-dimensional space. The downside of SVM is that it is computationally more expensive than other ML models. Despite that, the model is still widely used as it is less prone to overfitting. 
\subsubsection{Naïve Bayes}
The Naïve Bayes (NB) classifier \citep{lewis1998naive} is a simple probabilistic classifier based on the Bayes theorem. The equation of the Bayes theorem is shown in Eq.~\eqref{eq:1}. 
\begin{equation}
P(A \mid B) = \frac{P(B \mid A) P(A)}{P(B)}
\label{eq:1}
\end{equation}
From Bayes theorem, it is given that the probability of $A$ given $B$ is equal to the probability of $B$ given $A$ times the probability of $A$ and divided by the probability of $B$ \citep{lewis1998naive}. In other words, we can calculate $P(A \mid B)$ if $P(B \mid A)$, $P(A)$ and $P(B)$ is known. Therefore, the theorem can calculate the probability of data being in each class based on the given information and classify the data. In the training stage, the necessary probabilities are computed using the training data. As probability calculation is a simple task, the model is very efficient. Although the model is “naïve” as it assumes that each feature is independent of the other, it sometimes achieves comparable accuracy to other models.
\subsubsection{Artificial Neural Network}
\begin{figure}[!htbp]
\centering
\includegraphics[width=0.3\textwidth]{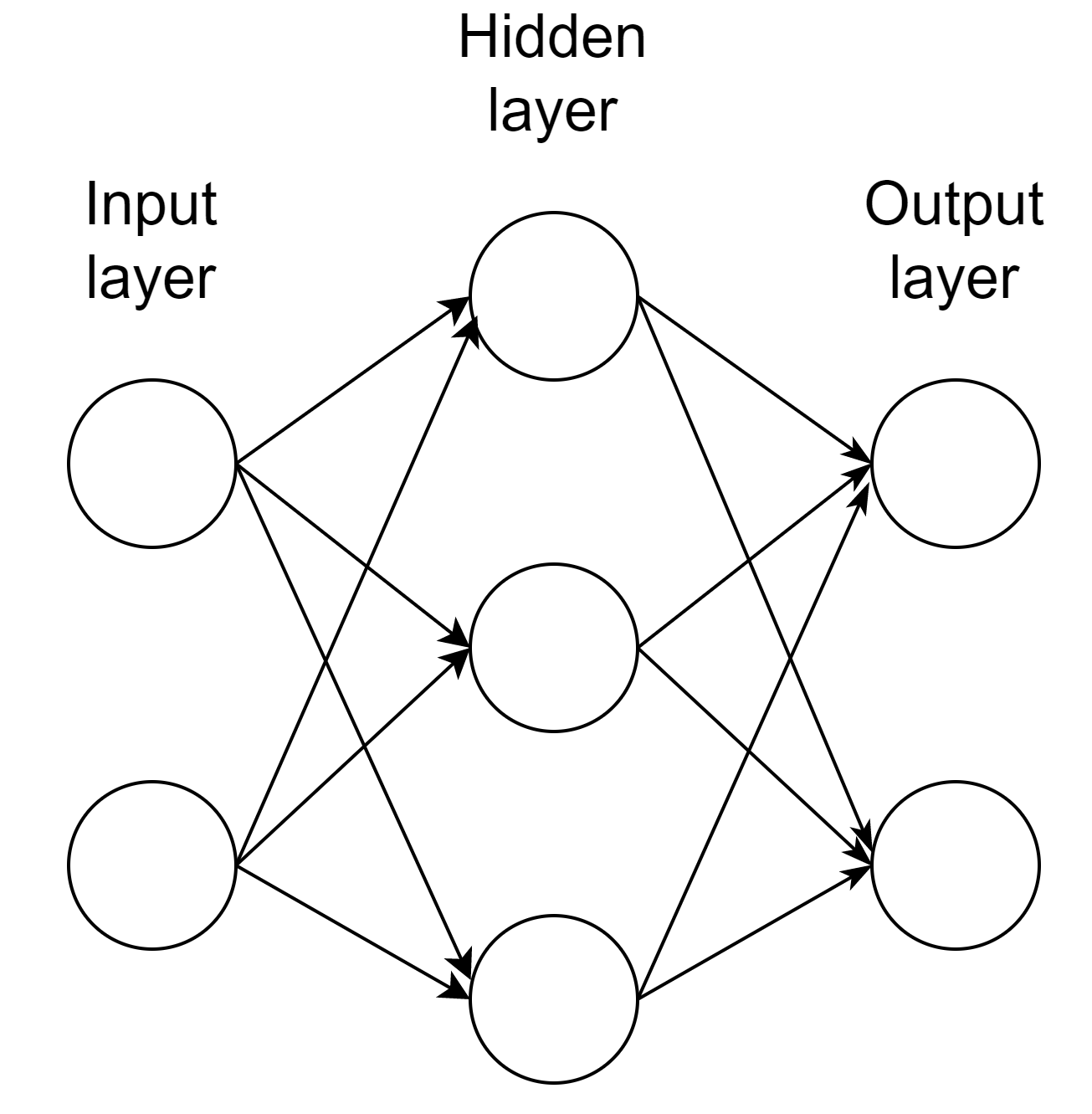}
\caption{A simple MLP with one hidden layer.}
\label{fig:5}
\end{figure}

Artificial neural network (ANN) is a more complicated model than the models discussed above. The multi-layer perceptron (MLP), a type of ANN, is adapted in this work. The structure of an MLP is typically represented using a graph as shown in Figure~\ref{fig:5}. As its name suggests, an MLP consists of multiple layers of perceptron. There are multiple nodes on each perceptron layer, where the nodes are also referred to as neurons. The example shown in Figure~\ref{fig:5} is an MLP consisting of three perceptron layers. 
The leftmost layer is also called the input layer. Each neuron from the input layer represents one feature of the input data. The rightmost layer is known as the output layer, and each neuron to the output layer represent the class of the data \citep{liu2017survey}. The neurons between different layers are connected by weighted vertices.  

\begin{figure}[!htbp]
\centering
\includegraphics[width=0.3\textwidth]{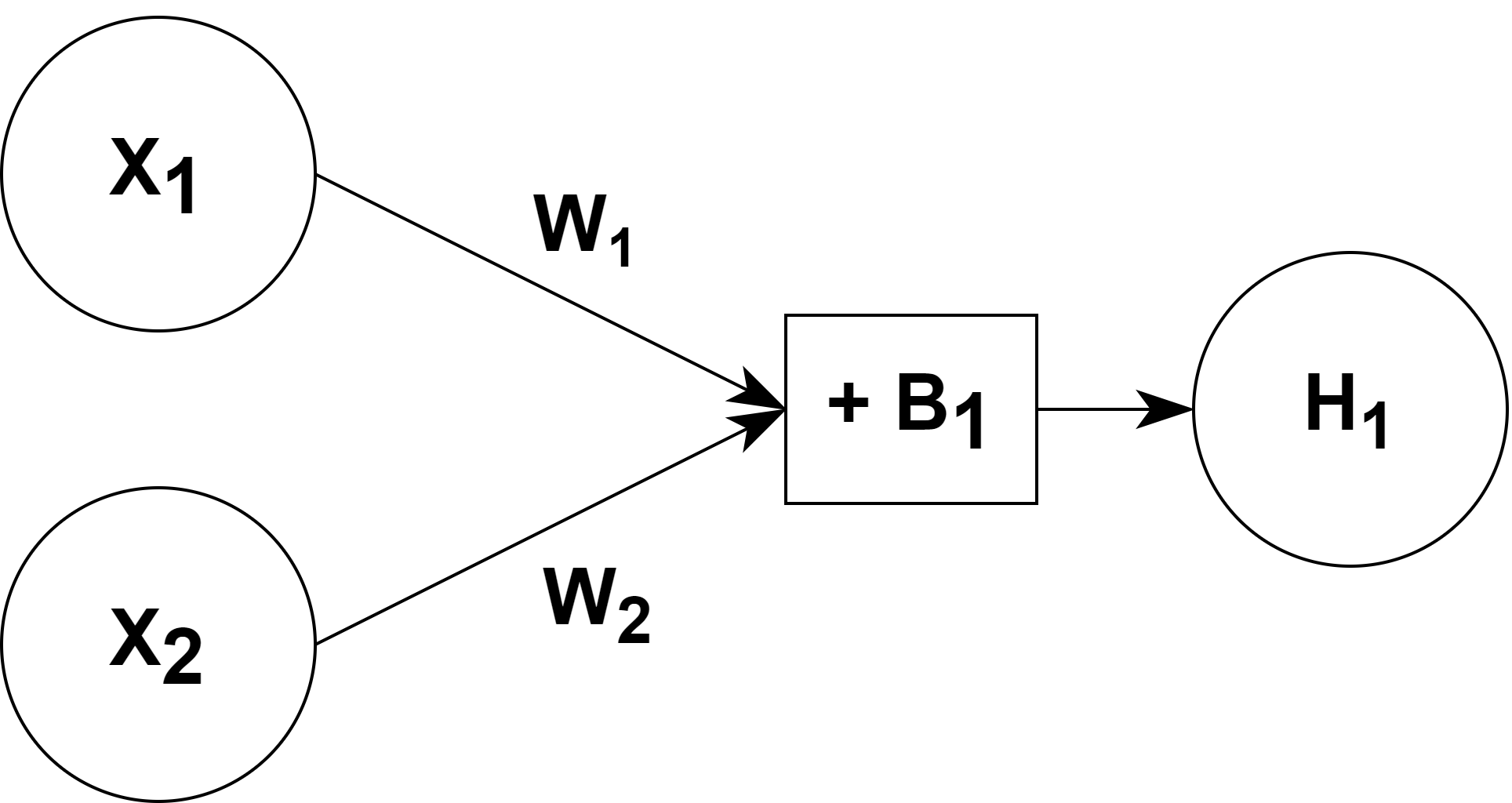}
\caption{Mapping to a neuron on next perceptron.}
\label{fig:6}
\end{figure}

The mapping from a perceptron to a neuron on the next layer is as simple as a linear function and a non-linear activation function. The mapping illustrated in Figure~\ref{fig:6} can be represented with the following Equation \citep{vinayakumar2019deep,bengio2009learning}:
\begin{equation}
h_{i}(x) = f(w_{i}^{T}x + b),
\end{equation}
where $x$ denotes the inputs to the next layer, $w_i$ represents the weights on the vertices, $b$ denotes the biases, and $f$ denotes the activation function. When training an ANN, the task is to optimise the weights $(w)$ and biases $(b)$ so that the model will fit the training data. As an ANN often includes a large number of neurons, there will be many parameters to be optimised. Hence, training an ANN is computationally more expensive than other models like decision trees and naïve Bayes. Besides that, a large dataset is required to train the neural network to prevent overfitting \citep{liu2017survey}. 

The main advantage of neural networks is that it overcomes the limitation of processing the data in its raw form. 
Conventional models like decision trees are limited when processing the raw data; hence, they heavily rely on hand-designed features. In other words, neural networks have a better chance to achieve good accuracy when the features of the training dataset are not well optimised. The neural networks overcome the limitation by transforming the data to a more abstract representation as the data go through different perceptron. As a result, neural networks can learn very complex patterns \citep{lecun2015deep}.
\subsubsection{Deep Neural Network}
The Deep Neural Network (DNN) has gained popularity since Hinton, Osindero, and Teh \citep{hinton2006fast} had successfully trained a DNN with three hidden layers in 2006. The application of DNN in the domain of IDS has also gained popularity in recent years \citep{ferrag2020deep}. The DNN is classified as deep learning (DL) model, which is an MLP with multiple hidden layers \citep{bengio2009learning}. Compared to MLP with a single hidden layer, DNN performs better when pretraining is eliminated \citep{hinton2006reducing}. Besides that, DNN has the ability to learn more complicated patterns with its deep architecture. 
With the additional complexity, DNNs take longer to train and require a larger dataset to optimise its parameter.
\subsection{Feature Selection}
\label{sec:feature_selection}
Besides using the suitable algorithm and optimising its hyperparameters, feature selection is also an important aspect to improve the performance of an IDS~\citep{aksu2018intrusion}. The feature selection aims to select a concise set of features to classify the network traffic. A modern dataset like the CSE-CIC-IDS2018 may contain around 80 features; such a high dimensional dataset significantly increases the complexity and duration for training the model. By reducing the number of features, not only the training time can be reduced, but the accuracy of the model can also be improved since unrelated information has been filtered out.
 
Random forest is one of the most widely used algorithms to perform feature selection.  
Li et al. \citep{li2020building} conducted a study to improve the accuracy of the neural network while reducing the training time by reducing the number of features. A random forest is built to reduce the number of features, and the features are ranked according to their respective permutation importance score \citep{gregorutti2017correlation} calculated during the training process. The most important features are then selected and grouped into an even smaller set of features. The proposed model was compared to KitNET \citep{mirsky2018kitsune} by using the CSE-CIC-IDS2018 dataset. In the paper, Li et al. show that the number of features can be reduced from 80 to around 20. Compared to KitNET, the authors show that the model has comparable accuracy with KitNET; at the same time, the detection time was significantly reduced.
\section{Related works}
\label{sec:related_works}
The adoption of ML in IDS has been widely explored over the last decade. A review by García-Teodoro et al. \citep{garcia2009anomaly} shows that researchers had been working in the domain since the 2000s. Researchers are still developing more advanced algorithms to improve the accuracy and efficiency of the ML-based IDS.

In 2019, Kaja, Shaout and Ma \citep{kaja2019intelligent} proposed a two-stage IDS that combines an unsupervised model and a supervised model to mitigate the false positive and false negative. In the first stage, the model uses K-Means Clustering, an unsupervised model, to detect malicious activity. In the second stage, supervised models like Decision Tree, Random Forest and Naïve Bayes are used to classify the malicious activity. The authors showed that the proposed models could achieve 99.97\% to 92.74\% accuracy on the KDD99 dataset. In the same year, Kanimozhi and Jacob \citep{kanimozhi2019artificial} conducted a study on utilising ANN for IDS. Their study focused on detecting the botnet attacks in the CSE-CIC-IDS2018 dataset. With hyperparameter optimisation, they achieved an accuracy of 99.97\%, with an average false positive rate of only 0.001. 

Besides conventional ML models like SVM, Decision Tree and Naïve Bayes, researchers are also exploring the application of modern models like Deep Learning in the domain of IDS. The review by Ferrag et al. \citep{ferrag2020deep} shows that more than 40 works have been done for the adoption of Deep Learning in the domain of IDS. In 2019, Vinayakumar et al. \citep{vinayakumar2019deep} conducted a study focusing on the adoption of Deep Learning in both HIDS and NIDS. The study proposed a scalable and hybrid DNNs framework. The authors showed that the proposed DNNs framework could achieve an accuracy rate of 93.5\%. Although the accuracy rate is not as high compared to DT and RF, the proposed framework is claimed to be computationally inexpensive for training. 

We can see from the recent literature that IDS with ML models can easily achieve a higher than 90\% of accuracy. Hence, some literature focuses on a comparative study, where multiple models are implemented and evaluated using various datasets. In 2018, Verma and Ranga \citep{ranga2018evaluation} implemented ten models for IDS, including both supervised and unsupervised ML models. The authors evaluated multiple ML models, including ANN, Deep Learning, KNN and SVM. They provided a detailed comparative study using the CIDDS-001 dataset. The result shows that KNN, SVM, DT, RF and DL are the best performing models, with accuracy rates above 99.9\%.
 
In 2020, Ferrag et al. \citep{ferrag2020deep} summarised more than 40 works that implemented Deep Learning for IDS and described 35 well-known datasets in the domain of IDS. The authors also implemented seven deep learning models and compared the performance with Naïve Bayes, ANN, SVM and Random Forest. In the study, they used CSE-CIC-IDS2018 and the Bot-IoT datasets. The result shows that the Deep Learning models achieved a 95\% of detection rate, which outperformed the 90\% of detection rate achieved by other models. 

In 2021, a comprehensive literature review was conducted by Kilincer et al. \citep{kilincer2021machine} 
to compare the performance of SVM, KNN and DT. Multiple datasets were used for the comparative analysis, including the CSE-CIC-IDS2018, UNSW-NB15, ISCX-2012, NSL-KDD and CIDDS-001 datasets. The result from the study shows that the accuracy of the models ranged from 95\% to 100\% except for the UNSW-NB15 dataset. DT consistently performed the best among all implemented models regardless of the dataset.

When evaluating the performance of the IDS, the ability to detect unknown attacks is also an area of concern. In 2020, Hindy et al. \citep{hindy2020utilising} focused on the performance of ML-based IDS on detecting unknown attacks. The study proposed an IDS to detect zero-day attacks with high recall rates while keeping the miss rate to a minimum. Besides that, they implemented a one-class SVM to compare with the proposed model. The study used the CIC-IDS2017 dataset and the NSL-KDD dataset for model training and evaluation. To fulfil the setting of zero-day attack detection, only the normal traffics were used when training the model and all attack activities were used to mimic zero-day attacks. The result from the study showed that both models had a low miss rate on detecting zero-day attacks.

As summarised in Table~\ref{tab:1}, we can see that recent studies commonly involve multiple ML models and multiple datasets for evaluation. However, the evaluation using different datasets is done separately. The models are retrained when evaluating the dataset using a different dataset. This is because different dataset has different feature set. In this paper, we identified multiple pairs of datasets for training and testing. Each pair of datasets share the same feature set. Besides that, in our work, the testing dataset is created later than the training dataset so that the long-term performance of IDS can be better reflected.
\begin{table}[!htbp]
\centering
\caption{ML models and datasets that used by recent literature.}
\label{tab:1}
\begin{tabular}{c|c|c}
\hline
Ref. & ML models & Datasets used \\
\thickhline

\citep{ranga2018evaluation} & 
\makecell{ANN, DL, RF, KNN, SVM, DT, NB,\\ KMC, EMC, SOM} & 	CIDDS-001 \\
\hline

\citep{kaja2019intelligent} &
KMC + DT, RF, NB	&KDD99
\\ \hline

\citep{vinayakumar2019deep} & \makecell{DNNs, RF, KNN, SVM, DT, NB} & \makecell{KDD99, NSL-KDD, UNSW-NB15,\\ Kyoto, WSN-DS, CIC-IDS2017}
\\ \hline

\citep{kanimozhi2019artificial} & ANN	& CSE-CIC-IDS2018
\\ \hline

\citep{ferrag2020deep} &	DNNs, NB, ANN, SVM, RF	& CSE-CIC-IDS2018, Bot-IoT
\\ \hline

\citep{hindy2020utilising} & ANN, SVM	& CIC-IDS2017, NSL-KDD
\\ \hline

\citep{li2020building} & DNN	& CSE-CIC-IDS2018
\\ \hline

\citep{kilincer2021machine} & SVM, KNN, DT &	CSE-CIC-IDS2018, NSL-KDD, CIDDS-001, ISCX2012
\\ \hline
\end{tabular}
\end{table}
\section{Framework of Experiment}
\label{sec:framework_of_experiment}
\begin{figure}[!htbp]
\centering
\includegraphics[width=0.4\textwidth]{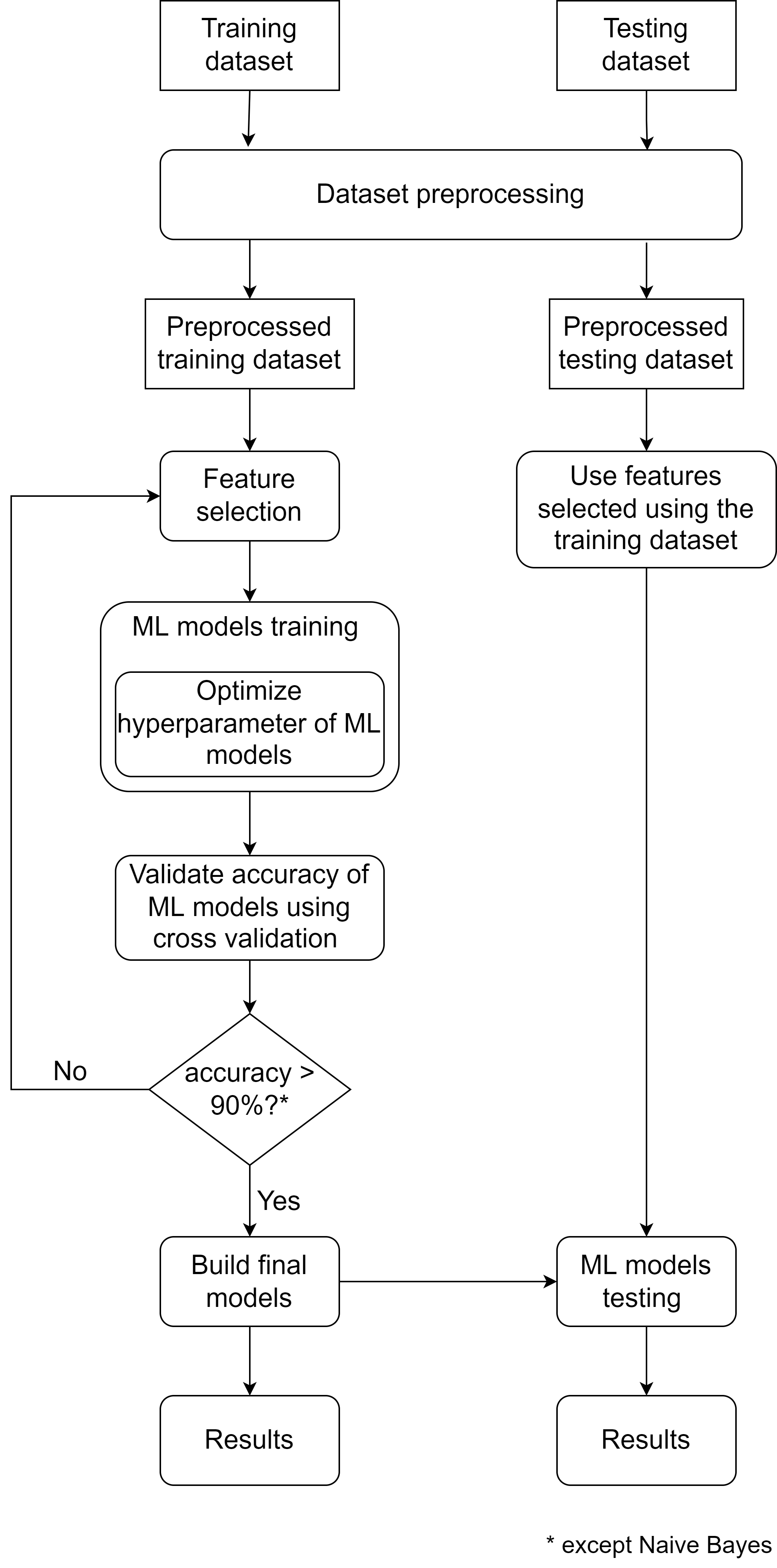}
\caption{The workflow of the experiment process.}
\label{fig:7}
\end{figure}

The experiments of this paper are separated into five main steps. The first step is dataset pre-processing, where the datasets are cleaned, and the necessary processing is carried out. The second step is feature selection, which is an important step to improve the performance of the models as described in Section~\ref{sec:feature_selection}. The third step is to optimise the hyperparameters to improve the accuracy of the models. The accuracy of the models is validated using cross-validation after optimising the hyperparameters. Finally, the performances of the models on the testing dataset are evaluated using various metrics. The complete flow of the experiment is illustrated in Figure~\ref{fig:7}. The details of each experiment step are further discussed in Section~\ref{sec:data_pre_processing} to Section~\ref{sec:model_evaluation}.

The experiment of this paper is conducted twice, each using a different set of datasets. The first set of datasets is the CIC-IDS2017 dataset \citep{sharafaldin2018toward} and the CSE-CIC-IDS2018 dataset. As described in Section~\ref{sec:dataset}, both datasets are created by Canadian Institute for Cybersecurity (CIC). As created by the same organisation, both datasets share a common set of features. Besides that, the CSE-CIC-IDS2018 dataset was created one year after the CIC-IDS2017 dataset. Hence, these datasets meet the needs of this paper, and the CIC-IDS2017 dataset is used as the training dataset while the CSE-CIC-IDS2018 dataset is used as the testing dataset. 

The second set of datasets is derived from the LUFlow dataset \citep{mills2021practical}. Although it may seem contradictory to the aim of our experiment, it is important to note that the LUFlow dataset organises the data according to the day that the data is collected. In this experiment, the data collected in July 2020 is used as the training dataset, while the data collected in January 2021 is used as the testing dataset.
\subsection{Data Pre-Processing}
\label{sec:data_pre_processing}
The first step of the experiment is to pre-process the dataset. First, we perform dataset cleaning by eliminating unwanted entries in the dataset. Entries containing missing or infinity values are dropped as they only contribute to a relatively small portion of the dataset. Besides that, we also remove duplicates to expose the models to as many unique samples as possible. 

Next, we address the high-class imbalance problem, where some classes have significantly more samples than others. The problem results in a bias towards the majority class, which in turn makes the accuracy of the model meaningless. We downsample the majority class to address the high-class imbalance problem. In our approach, samples are selected randomly to reduce the number of samples for the majority class. Besides that, multiple minority classes are combined to form a larger class. 

We also ensure that both datasets have the same set of columns with the same sequence, as we proposed using different datasets for training and testing. Besides that, both datasets are relabelled, if necessary, to ensure that both datasets have the same classes.
\subsection{Feature Selection}
It is important to perform feature selection before training the model. Since modern datasets like the CIC-IDS2017 dataset contain around 80 features, training the model without any feature selection will consume time. Other than that, some features may include noise and reduce the accuracy of the model. As pointed out by Aksu et al. \citep{aksu2018intrusion}, the accuracy of the models starts to drop when more than 40 features of the CIC-IDS2017 dataset are used to train the model.

We use the random forest algorithm by utilising the RandomForestClassifier provided by scikit-learn to perform feature selection. A random forest is trained using the training data, and the top $n$ features with the highest importance score are selected. The reason for choosing random forest is because it is widely used in the domain of IDS. As an example, Sharafaldin, Lashkari and Ghorbani \citep{sharafaldin2018toward} and Kostas \citep{kostas2018anomaly} also used the random forest for feature selection on the CIC-IDS2017 dataset.

After reducing the number of features using random forest, we further reduce the number of features using brute force. We then built preliminary ML models using a different number of features. A for loop is used to add a new feature in each iteration to construct the ML models until all $n$ features are included. The accuracy of each model is recorded with respect to the number of features. Based on the accuracy rate of the models, a more concise set of features is selected. 

Moreover, some features are removed by human inspection. Some features should be removed although having a high correlation with the output variable, source IP address, for example, is one of them. When a dataset includes a large amount of malicious traffic from one IP address, the “source IP address” may be ranked as an important feature by the random forest. However, the feature should be removed to prevent overfitting, as classifying the traffic based on the IP address may not be relevant in the future. 
\subsection{ML Models Training}
We train the models using the training dataset after selecting the features. When training the models, the hyperparameters of the models are optimised using grid search by utilising the GridSeachCV function provided by scikit-learn. The grid search method works by searching through a predefined hyperparameter space, and different combinations of hyperparameters are used to train the model. The hyperparameters that give the best accuracy is chosen to train the final models. The hyperparameters of a model are parameters that govern the training process. Take DNN as an example; the number of hidden layers is the hyperparameter, while the weights and biases of the neural network are the parameters of the model. Since optimising the hyperparameters require training the models multiple times, only a fraction of the training dataset is used to speed up the entire process. 
\subsection{Model Accuracy Verification}
\begin{figure}[!htbp]
\centering
\includegraphics[width=0.5\textwidth]{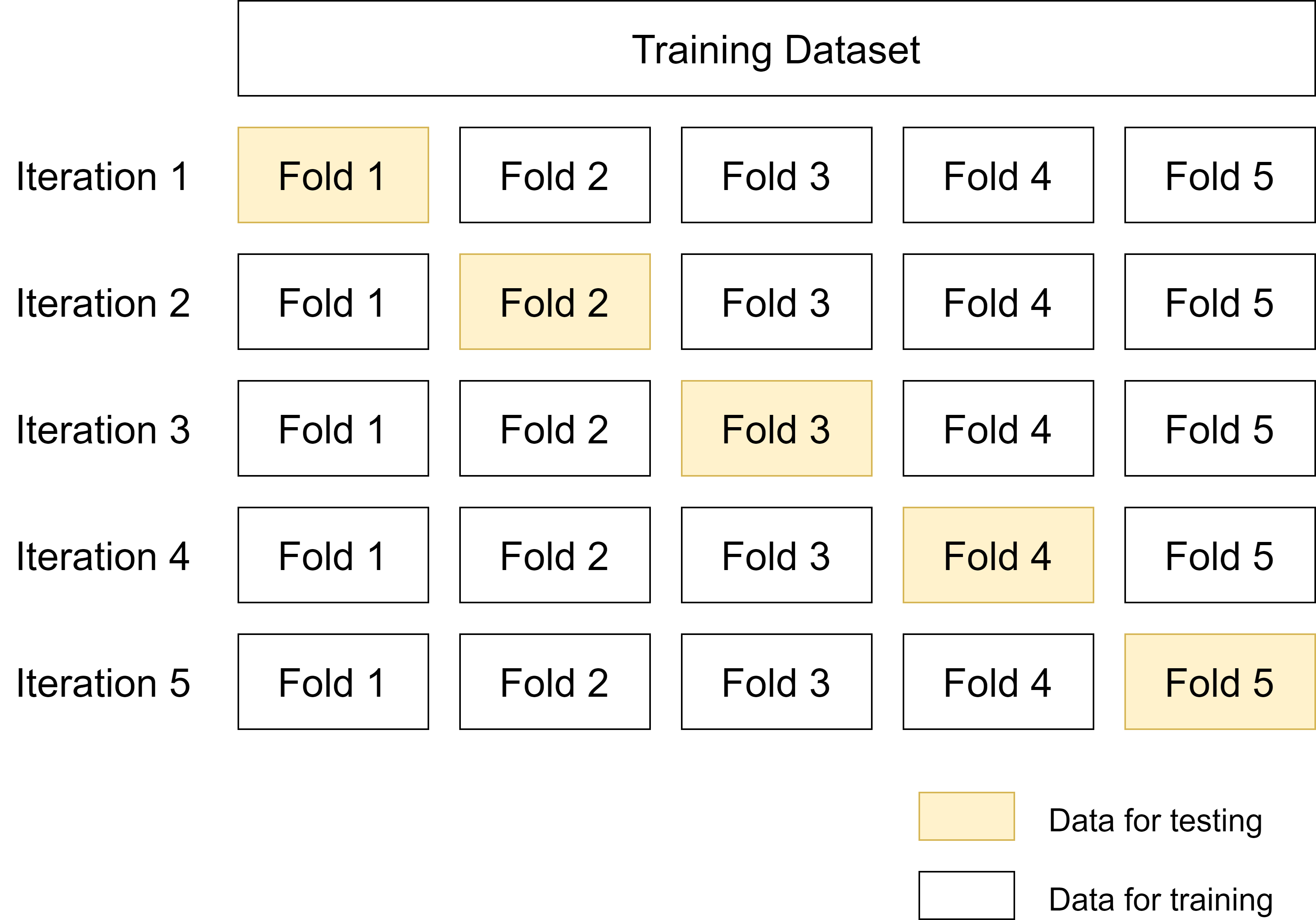}
\caption{An example of $k$ fold cross-validation with $k=5$.}
\label{fig:8}
\end{figure}

In this step, the goal is to verify the accuracy of the models. Before we test the models using a different dataset, it is important to ensure that they perform well on the training dataset. Hence, we measure the performance of the models using cross-validation. The training dataset is split into $k$-folds, and the models are validated through $k$ iteration. In each iteration $i$, the $i$-th fold of the training dataset is used for testing, and the rest of the training dataset is used for training (as illustrated in Figure~\ref{fig:8}). The average accuracy through the cross-validation is the accuracy of the model. After that, the accuracy of the models is compared with other literature. If the model achieved a comparable accuracy with other literature, we would proceed to the next step. Otherwise, the experiment process is revised for improvement.
\subsection{Model Evaluation}
\label{sec:model_evaluation}
After verifying the accuracy of the models, the last and most crucial step of this paper is performed. The last step is training the final model and testing the models using the testing dataset. First, we use 70\% of the training dataset to train the final models. Next, we use the rest of the training dataset to calculate the accuracy of the models on the training dataset. Finally, we use the testing dataset to test the models. 

The interest of this paper is to compare the accuracy of each model when a different dataset is being used. At the same time, a comparison between different models is conducted in terms of accuracy and efficiency. The performance metrics that are used to measure the performance of the models include accuracy, precision, recall and F1-score, as shown in Eq.~\eqref{eq:3}, Eq.\eqref{eq:4}, Eq.\eqref{eq:5} and Eq.\eqref{eq:6}.
\begin{align}
Accuracy = \frac{TP+TN}{TP+TN+FP+FN} \label{eq:3}\\
Precision = \frac{TP}{TP+FP} \label{eq:4}\\
Recall = \frac{TP}{TP+FN} \label{eq:5}\\
F1 Score = 2 \times \frac{Precision \times Recall}{Precision + Recall} \label{eq:6}
\end{align}

In the above equations, true positive (TP) and true negative (TN) denote the number of samples that are correctly classified as positive and negative, respectively. False positive (FP) and false negative (FN) indicate the number of incorrectly classified samples as positive and negative, respectively. Besides that, we also visualise the classification result using a confusion matrix.

Moreover, we also measure the time complexity of each model. Our primary focus is the time consumption of each model to be trained and the time consumption for prediction. We do not compare with other literature regarding time efficiency as the time consumption depends on the implementation of the model, the number of samples and the hardware used to execute the experiment.

\section{Experiments and Discussions}
\label{sec:experiment_and_discussion}
In this section, the experiment results are discussed in detail. Section~\ref{sec:exp_environment} discusses the software and hardware environments used for our experiments. Section~\ref{sec:exp_cic_dataset} discusses the experiment results using the CIC-IDS2017 and the CSE-CIC-IDS2018 datasets, while Section~\ref{sec:exp_luflow_dataset} discusses the experiment results using the LUFlow dataset. The experiment details of this paper are also made publicly accessible via a GitHub repository; see reference \citep{web:tuanhong} to access the source codes for all our experiments.
\subsection{Experiment Environment} 
\label{sec:exp_environment}
We first introduce the experiment environment of this work. The environment for the experiment is important as it will affect the performance of the implementation. 
The ML models of this paper are implemented using Python programming language in conda environment. The conda version is 4.10.1, which use Python 3.8.8. Python libraries, including scikit-learn \citep{pedregosa2011scikit}, pandas, NumPy and Matplotlib are being used for the implementation of this work. 
The hardware environment used for this paper is a laptop powered by the Intel® Core™ i7-8550U Central Processing Unit (CPU). The main frequency of the processor is 1.80GHz, equipped with 8GB of Random-Access Memory (RAM). The Graphics Processing Unit (GPU) of the system is Intel® UHD Graphics 620 integrated graphics unit. Besides that, the system is running on Windows 11 Home 64-bit operating system.

\subsection{Experiment Using CIC’s Dataset}
\label{sec:exp_cic_dataset}
In this section, the experiment results of using the datasets created by the CIC is being discussed. The CIC-IDS2017 dataset is used for training, and the CSE-CIC-IDS2018 dataset is used for testing.
\subsubsection{Dataset Pre-Processing}
Both the CIC-IDS2017 and the CSE-CIC-IDS2018 datasets are huge datasets, where the former contains 800MB of data and the latter contains 6.4GB of data. The CSE-CIC-IDS2018 dataset is too large to be handled by the hardware used in our work; hence, we only use 10\% of the dataset. 

Although the CIC-IDS2017 and CSE-CIC-IDS2018 are some of the most recently created datasets, they still require some pre-processing. In both datasets, there are a small number of entries that contain missing values or infinity values. Those entries are removed as they only contribute to a small portion of the dataset. Besides that, duplicates in both datasets are also removed. 

After the datasets are cleaned, the datasets are resampled to reduce the class imbalance problem. As shown in Table~\ref{tab:3} and Table~\ref{tab:4}, both datasets have severe class imbalance problems.
\begin{table}[!tb]
\centering
\caption{Class distribution of the CIC-IDS2017 dataset}
\label{tab:3}
\begin{tabular}{c|cc|cc}
\hline
\multirow{2}{*}{Classes} & \multicolumn{2}{c|}{Before cleaning} & \multicolumn{2}{c}{After cleaning and resampling}\\ 
 & No. of rows & No. of rows (\%) & No. of rows & No. of rows (\%) \\ \thickhline 
BENIGN &2273097 &80.3004\%	&324881	&50.0000\% \\ \hline
DoS Hulk	&231073	&8.1629\%	&100000	&15.3903\% \\ \hline
PortScan	&158930	&5.6144\%	&90694	&13.9580\% \\ \hline
DDoS	&128027	&4.5227\%	&100000	&15.3903\% \\ \hline
DoS GoldenEye	&10293	&0.3636\%	&10286	&1.5830\% \\ \hline
FTP-Patator	&7938	&0.2804\%	&5931	&0.9128\% \\ \hline
SSH-Patator	&5897	&0.2083\%	&3219	&0.4954\% \\ \hline
DoS slowloris	&5796	&0.2048\%	&5385	&0.8288\% \\ \hline
DoS Slowhttptest	&5499	&0.1943\%	&5228	&0.8046\% \\ \hline
Bot	&1966	&0.0695\%	&1948	&0.2998\% \\ \hline
Web Attack-Brute Force	&1507	&0.0532\%	&1470	&0.2262\% \\ \hline
Web Attack-XSS	&652	&0.0230\%	&652	&0.1003\% \\ \hline
Infiltration	&36	&0.0013\%	&36	&0.0055\% \\ \hline
Web Attack-Sql Injection	&21	&0.0007\%	&21	&0.0032\% \\ \hline
Heartbleed	&11	&0.0004\%	&11	&0.0017\% \\ \hline
\end{tabular}
\end{table}
\begin{table}[!tb]
\centering
\caption{Class distribution of the CSE-CIC-IDS2018 dataset (10\% of the entire dataset)}
\label{tab:4}
\begin{tabular}{c|cc|cc}
\hline
& \multicolumn{2}{c|}{Before cleaning} & \multicolumn{2}{c}{After cleaning and resampling}\\ 
Classes & No. of rows (10\%) & No. of rows (\%) & No. of rows & No. of rows (\%) \\ \thickhline 
Benign	&1347953	&83.0747\%	&257791	&50.0000\% \\ \hline
DDOS attack-HOIC	&68801	&4.2402\%	&68628	&13.3108\% \\ \hline
DDoS attacks-LOIC-HTTP	&57550	&3.5468\%	&57550	&11.1621\% \\ \hline
DoS attacks-Hulk	&46014	&2.8359\%	&45691	&8.8620\% \\ \hline
Bot	&28539	&1.7589\%	&28501	&5.5279\% \\ \hline
FTP-BruteForce	&19484	&1.2008\%	&12368	&2.3988\% \\ \hline
SSH-Bruteforce	&18485	&1.1392\%	&16312	&3.1638\% \\ \hline
Infilteration	&16160	&0.9959\%	&16034	&3.1099\% \\ \hline
DoS attacks-SlowHTTPTest	&14110	&0.8696\%	&7251	&1.4064\% \\ \hline
DoS attacks-GoldenEye	&4154	&0.2560\%	&4153	&0.8055\% \\ \hline
DoS attacks-Slowloris	&1076	&0.0663\%	&1049	&0.2035\% \\ \hline
DDOS attack-LOIC-UDP	&163	&0.0100\%	&163	&0.0316\% \\ \hline
Brute Force -Web	&59	&0.0036\%	&59	&0.0114\% \\ \hline
Brute Force -XSS	&25	&0.0015\%	&25	&0.0048\% \\ \hline
SQL Injection	&7	&0.0004\%	&7	&0.0014\% \\ \hline
\end{tabular}
\end{table}
The ratio of benign samples to malicious samples is around 4:1. Besides that, most malicious samples contribute to less than 1\% of the dataset. The benign class and some attack classes are downsampled to make the class distribution more balanced. For the CIC-IDS2017 dataset, attack classes containing more than 100000 samples are downsampled. After that, the benign class is also downsampled so that the ratio of benign samples to malicious samples is 1:1. 

For the CSE-CIC-IDS2018 dataset, the process is very similar, except that all malicious samples are used. The class distribution of both datasets after cleaning and resampling is shown in Table~\ref{tab:3} and Table~\ref{tab:4}, respectively. Although the class distribution is still uneven after resampling, minority classes are not upsampled as it will cause biases in the models. Instead, the malicious samples of both datasets are relabelled as ‘malicious’ to prevent the models overfit a certain attack class. 
\subsubsection{Feature Selection}
The feature selection is an important process when training the models using the CIC-IDS2017 dataset, as it includes 78 features. For the CIC dataset experiment, only the CIC-IDS2017 dataset is used for feature selection. We do not inspect the features manually as the dataset includes many features.  

\begin{figure}[!tb]
\centering
\includegraphics[width=\textwidth]{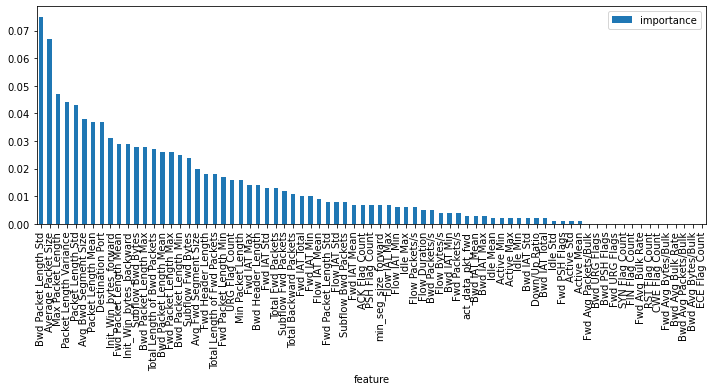}
\caption{Importance score of each feature in CIC-IDS2017 dataset.}
\label{fig:10}
\end{figure}

When performing feature selection using the random forest algorithm, 10\% of the dataset is used to train the model. The features are then ranked according to their respective importance scores calculated during the training process. The ranking of the features is displayed in Figure~\ref{fig:10}. 
As shown in Figure~\ref{fig:10}, the scores of the top two features are significantly higher than the rest. Other than that, most features have importance score of 0.03 to 0.01. From the result given by the random forest algorithm, the top 10 to 20 features are needed to train the models.

\begin{figure}[!htbp]
\centering
\includegraphics[width=0.9\textwidth]{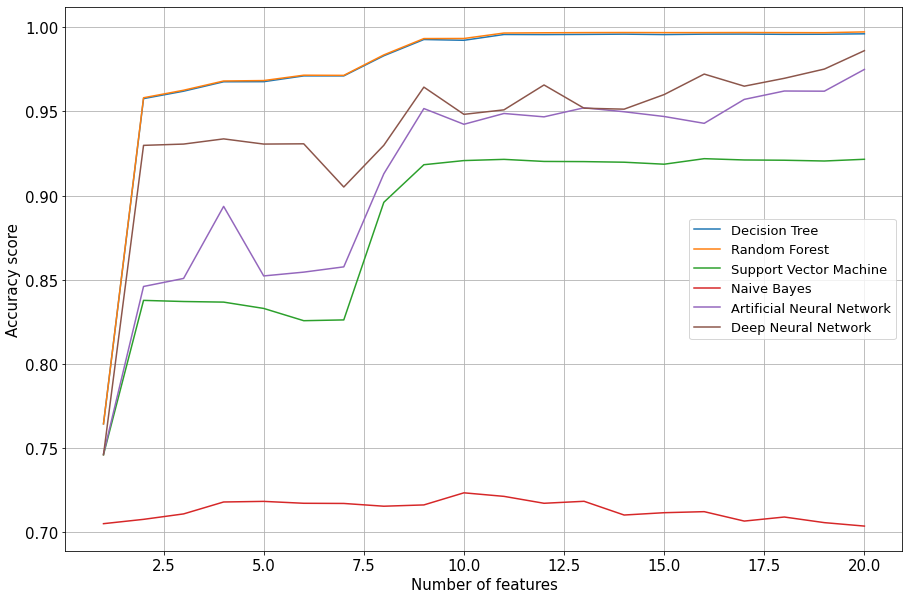}
\caption{Accuracy of the models with respect to the number of features.}
\label{fig:11}
\end{figure}

To further reduce the features, the brute force method is used. Preliminary ML models are trained by using the top-ranked features. Starting from Bwd Packet Length Std, one feature is added in each iteration according to their ranking. Figure~\ref{fig:11} shows the accuracy of the models as the number of features increases. From the figure, we can see that the top 11 features give the best overall accuracy. 
\subsubsection{Accuracy Validation}
In this step, the accuracy of the models trained using the optimal hyperparameter is validated. The accuracy score of each model is calculated using 5-fold cross-validation. The score of each model in each fold is displayed in Table~\ref{tab:5}. The table shows that all models achieved higher than 95\% of accuracy score except NB. Besides that, the standard deviation of the accuracy of each model is very minimal. As the maximum standard deviation is only 0.0029, the accuracy of each model on the CIC-IDS2017 dataset is very consistent. 
\begin{table}[!tb]
\centering
\caption{The accuracy result of 5-fold cross-validation on CIC-IDS2017 dataset.}
\label{tab:5}
\begin{tabular}{c|c|c|c|c}
\hline
Model & Fold & Accuracy & Mean accuracy & Standard deviation \\ \thickhline
\multirow{5}{*}{Decision Tree}&
Fold-1 & 0.9972 & \multirow{5}{*}{0.9970} & \multirow{5}{*}{0.0002} \\
& Fold-2 &0.9967 &  & \\ 
& Fold-3 &0.9971 & & \\
& Fold-4 &0.9971 & & \\ 
& Fold-5 &0.9970 & & \\ \hline
\multirow{5}{*}{Random Forest}&
Fold-1 & 0.9968 & \multirow{5}{*}{0.9972} & \multirow{5}{*}{0.0004} \\
& Fold-2 &0.9978 &  & \\ 
& Fold-3 &0.9974 & & \\
& Fold-4 &0.9968 & & \\ 
& Fold-5 &0.9971 & & \\ \hline
\multirow{5}{*}{Support Vector Machine}&
Fold-1 & 0.9654 & \multirow{5}{*}{0.9650} & \multirow{5}{*}{0.0008} \\
& Fold-2 &0.9641 &  & \\ 
& Fold-3 &0.9660 & & \\
& Fold-4 &0.9641 & & \\ 
& Fold-5 &0.9655 & & \\ \hline
\multirow{5}{*}{Naïve Bayes}&
Fold-1 & 0.7313 & \multirow{5}{*}{0.7311} & \multirow{5}{*}{0.0002} \\
& Fold-2 &0.7312 &  & \\ 
& Fold-3 &0.7307 & & \\
& Fold-4 &0.7311 & & \\ 
& Fold-5 &0.7313 & & \\ \hline
\multirow{5}{*}{Artificial Neural Network}&
Fold-1 & 0.9688 & \multirow{5}{*}{0.9693} & \multirow{5}{*}{0.0029} \\
& Fold-2 &0.9678 &  & \\ 
& Fold-3 &0.9705 & & \\
& Fold-4 &0.9741 & & \\ 
& Fold-5 &0.9653 & & \\ \hline
\multirow{5}{*}{Deep Neural Network}&
Fold-1 & 0.9773 & \multirow{5}{*}{0.9772} & \multirow{5}{*}{0.0006} \\
& Fold-2 &0.9777 &  & \\ 
& Fold-3 &0.9777 & & \\
& Fold-4 &0.9770 & & \\ 
& Fold-5 &0.9761 & & \\ \hline
\end{tabular}
\end{table}

On the other hand, the NB has the worst performance, with only 73\% accuracy. As the original target was to ensure all models at least achieve 90\% of accuracy, efforts have been made to improve the accuracy of NB. For example, the feature selection process has been improved by including the brute force method. Besides that, different variances of NB algorithms have been tested, including Gaussian Naïve Bayes and Bernoulli Naïve Bayes. However, the accuracy of NB did not improve much. 

As the accuracy of the model has been validated using cross-validation, the accuracy of each model is compared against two literature, including the work done by Kostas \citep{kostas2018anomaly} and the work done by Vinayakumar et al. \citep{vinayakumar2019deep}. The reason for choosing these two works is that both studies used the CIC-IDS2017 dataset, and most models used by this paper are included in their studies. Besides that, both works include binary classification, where the data are only classified as benign or malicious. Moreover, Kostas also performed feature selection using the random forest algorithm; hence, the work is comparable to ours.

Table~\ref{tab:6} shows that the models implemented in this paper are comparable to other models. The models using DT, RF, and SVM have a clear lead compared to the other existing works. The accuracy result of NB is between the two studies. The NB implemented by Kostas \citep{kostas2018anomaly} performed significantly better than that of this paper. In contrast, the NB implemented by Vinayakumar et al. \citep{vinayakumar2019deep} had much poorer accuracy than ours. 
\begin{table}[!tb]
\centering
\caption{Comparison of accuracy with other literature when using the CIC dataset.}
\label{tab:6}
\begin{tabular}{c|ccc}
\hline
\multirow{2}{*}{ML Models} & \multicolumn{3}{c}{Accuracy}\\ \cline{2-4}
& This work & Kostas \citep{kostas2018anomaly} & Vinayakumar et al. \citep{vinayakumar2019deep} \\ \thickhline
Decision Tree	&1.00	&0.95	&0.94 \\ \hline
Random Forest	&1.00	&0.94	&0.94 \\ \hline
Support Vector Machine	&0.96	&-	&0.80 \\ \hline
Naïve Bayes	&0.73	&0.87	&0.31 \\ \hline
Artificial Neural Network	&0.95	&0.97	&0.96 \\ \hline
Deep Neural Network	&0.97	&-	&0.94 \\ \hline

\end{tabular}
\end{table}

The difference is due to the fact that Kostas performed model-specific feature selection. In other words, the features used to train each model in Kostas’s work are different. With such optimisation, the NB is able to achieve significantly better accuracy. On the other hand, the accuracies of the models implemented by Vinayakumar et al. are lower than that of this work, except ANN. This is because the focus of Vinayakumar et al. is on proposing an optimised DNN that is less costly to train. Other models implemented by Vinayakumar et al. are just to provide a reference on how their proposed models perform. Hence, they may not perform optimisations to achieve the maximum possible accuracy. Besides that, it is important to note that the ANN and DNN of this paper are compared against the DNN proposed by Vinayakumar et al. with one and three hidden layers, respectively. 
Overall, the models implemented in this paper achieved a comparable or better accuracy than the other literature.
\subsubsection{Model Evaluation}
\begin{figure}[!htbp]
\centering
\includegraphics[width=\textwidth]{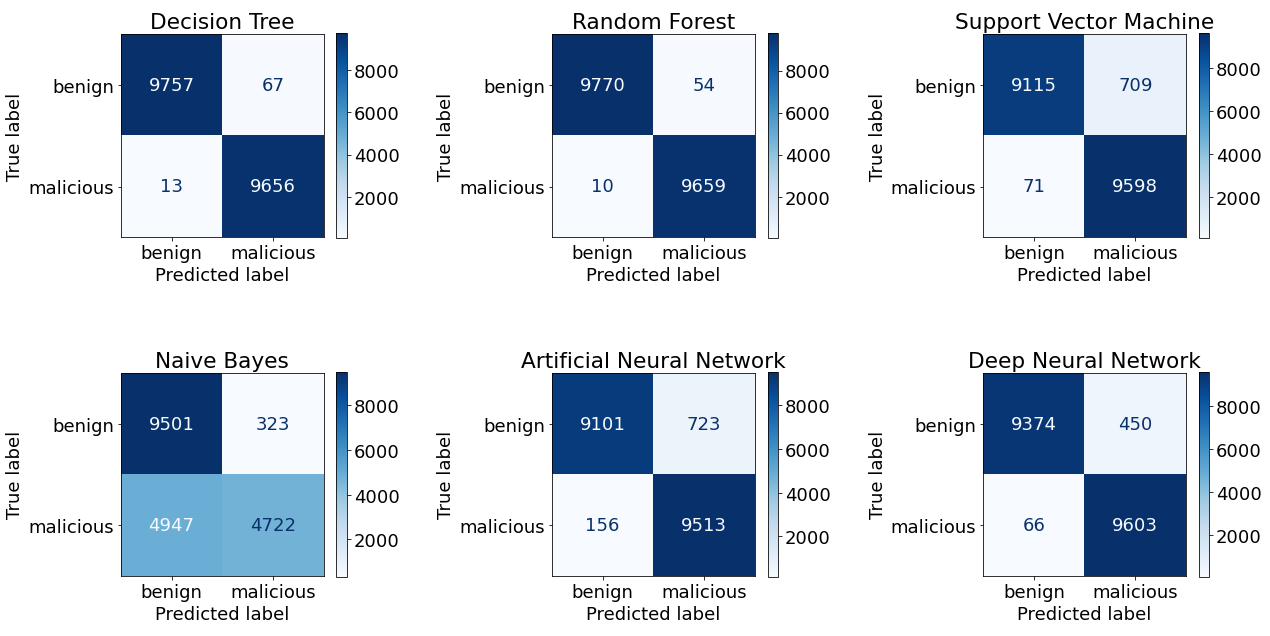}
\caption{Confusion matrix of each model on CIC-IDS2017 dataset.}
\label{fig:12}
\end{figure}

After ensuring that the models have achieved good accuracy, the models are tested using the testing dataset, the CSE-CIC-IDS2018 dataset. As described in Section~\ref{sec:model_evaluation}, the models are trained again using the optimal hyperparameters and tested again using the training dataset. This step is necessary to evaluate the performance of the models on the training dataset. The accuracy and the F1-score of the models on the training dataset are displayed in Table~\ref{tab:7}. Besides that, the accuracies of the models are also visualised using the confusion matrixes as shown in Figure~\ref{fig:12}. The confusion matrix clearly shows that DT and RF have the best accuracy, followed by SVM, ANN and DNN, while NB is biased towards benign class. 
\begin{table}[!htbp]
\centering
\caption{Performance of the models on the CIC-IDS2017 dataset.}
\label{tab:7}
\begin{tabular}{c|c|c|c|c|c}
\hline
\multirow{2}{*}{Models} & \multicolumn{4}{c|}{Evaluation Metrics} & \multirow{2}{*}{Class}\\ \cline{2-5}
& Accuracy & Precision & Recall & F1-score & \\ \thickhline
\multirow{2}{*}{Decision Tree} & \multirow{2}{*}{0.9959} & 0.9987 & 0.9932 & 0.9959 & benign \\
& & 0.9931 & 0.9987 & 0.9959 & malicious \\ \hline
\multirow{2}{*}{Random Forest} & \multirow{2}{*}{0.9967} & 0.9990 & 0.9945 & 0.9967 & benign \\
& & 0.9944 & 0.9990 & 0.9967 & malicious \\ \hline
\multirow{2}{*}{Support Vector Machine} & \multirow{2}{*}{0.9600} & 0.9923 & 0.9278 & 0.9590 & benign \\
& & 0.9312 & 0.9927 & 0.9610 & malicious \\ \hline
\multirow{2}{*}{Naïve Bayes} & \multirow{2}{*}{0.7296} & 0.6576 & 0.9671 & 0.7829 & benign \\
& & 0.9360 & 0.4884 & 0.6418 & malicious \\ \hline
\multirow{2}{*}{Artificial Neural Network} & \multirow{2}{*}{0.9549} & 0.9831 & 0.9264 & 0.9539 & benign \\
& & 0.9294 & 0.9839 & 0.9558 & malicious \\ \hline
\multirow{2}{*}{Deep Neural Network} & \multirow{2}{*}{0.9735} & 0.9930 & 0.9542 & 0.9732 & benign \\
& & 0.9552 & 0.9932 & 0.9738 & malicious \\ \hline
\end{tabular}
\end{table}

\begin{figure}[!tb]
\centering
\includegraphics[width=\textwidth]{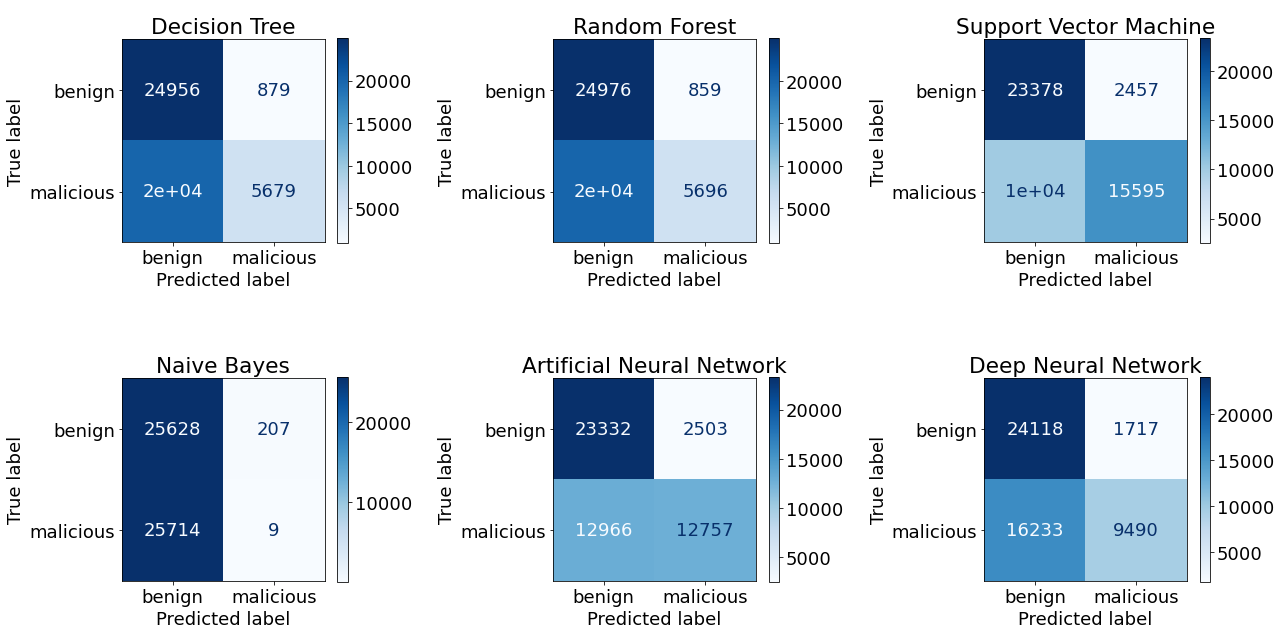}
\caption{Confusion matrix of each model on CSE-CIC-IDS2018 dataset.}
\label{fig:13}
\end{figure}

The most interesting part of this paper is to evaluate the models using another dataset. Table~\ref{tab:8} shows the performance of the models on the CSE-CIC-IDS2018 dataset. The accuracy of each model is also visualised using confusion matrixes as shown in Figure~\ref{fig:13}. It is important to note that the models are not trained using the CSE-CIC-IDS2018 dataset. The confusion matrixes show that all models have poor performance except SVM. Looking at the accuracy score and F1-score, SVM has the highest accuracy, with 76\% accuracy, followed by ANN with 70\% accuracy. The confusion matrixes also tell the problem of the models -- bias towards the benign samples. In other words, the models have overfitted the malicious samples to different extents. Among them, SVM suffers less from overfitting. From Figure~\ref{fig:13}, we can see that SVM allow more error, which incorrectly classifies more benign samples as malicious. In return, SVM correctly classifies more malicious samples.
\begin{table}[!tb]
\centering
\caption{Performance of the models on the CSE-CIC-IDS2018 dataset.}
\label{tab:8}
\begin{tabular}{c|c|c|c|c|c}
\hline
\multirow{2}{*}{Models} & \multicolumn{4}{c|}{Evaluation Metrics} & \multirow{2}{*}{Class}\\ \cline{2-5}
& Accuracy & Precision & Recall & F1-score & \\ \thickhline
\multirow{2}{*}{Decision Tree} & \multirow{2}{*}{0.5942} & 0.5546 & 0.9660 & 0.7046 & benign \\
& & 0.8660 & 0.2208 & 0.3518 & malicious \\ \hline
\multirow{2}{*}{Random Forest} & \multirow{2}{*}{0.5949} & 0.5550 & 0.9668 & 0.7052 & benign \\
& & 0.8690 & 0.2214 & 0.3529 & malicious \\ \hline
\multirow{2}{*}{Support Vector Machine} & \multirow{2}{*}{0.7559} & 0.6977 & 0.9049 & 0.7879 & benign \\
& & 0.8639 & 0.6063 & 0.7125 & malicious \\ \hline
\multirow{2}{*}{Naïve Bayes} & \multirow{2}{*}{0.4972} & 0.4992 & 0.9920 & 0.6641 & benign \\
& & 0.0417 & 0.0003 & 0.0007 & malicious \\ \hline
\multirow{2}{*}{Artificial Neural Network} & \multirow{2}{*}{0.7000} & 0.6428 & 0.9031 & 0.7510 & benign \\
& & 0.8360 & 0.4959 & 0.6226 & malicious \\ \hline
\multirow{2}{*}{Deep Neural Network} & \multirow{2}{*}{0.6518} & 0.5977 & 0.9335 & 0.7288 & benign \\
& & 0.8468 & 0.3689 & 0.5139 & malicious \\ \hline
\end{tabular}
\end{table}

\begin{figure}[!htbp]
    \centering
    \subfloat[\centering Accuracy of the models on the CIC’s dataset]{{\includegraphics[width=0.7\textwidth]{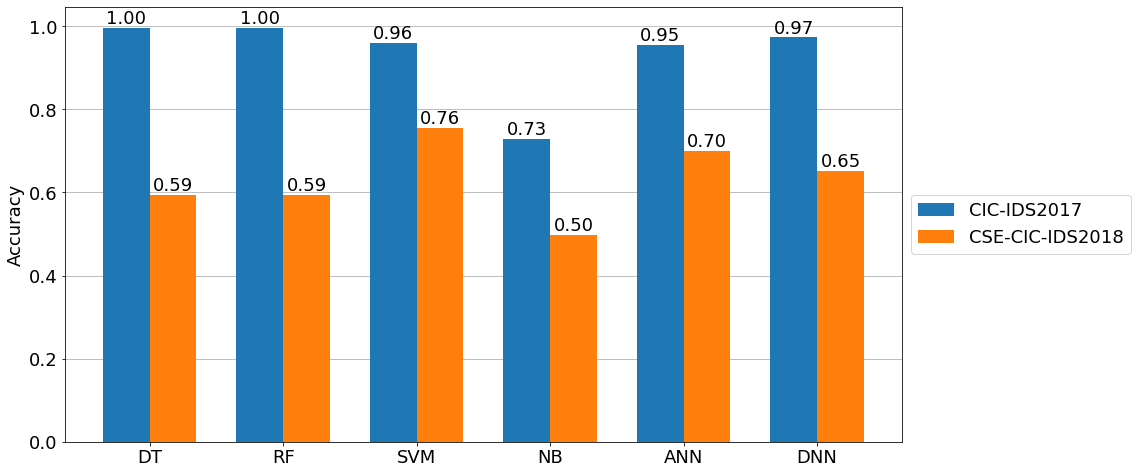} } \label{fig:14}}%
    \qquad
    \subfloat[\centering F1-score of the models on CIC’s dataset]
    {{\includegraphics[width=0.7\textwidth]{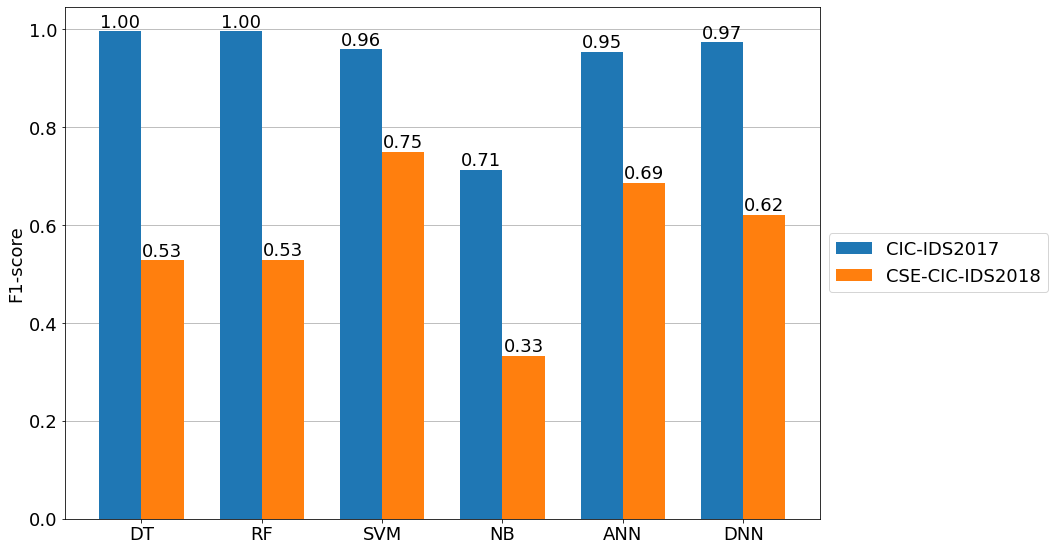} } \label{fig:15}}%
    \caption{Accuracy and F1-score of the models on the CIC’s dataset.}%
    \label{fig:14_&_15}%
\end{figure}

We use the data presented in Figure~\ref{fig:14} and Figure~\ref{fig:15} to compare which model suffers the most from overfitting. Figure~\ref{fig:14} compares the accuracy of the models on different datasets while Figure~\ref{fig:15} compares the F1-score. From Figure~\ref{fig:14}, we can see that SVM has the smallest drop in terms of accuracy, followed by ANN and NB. However, it is important to note that NB has a significant drop in terms of F1-score, as shown in Figure~\ref{fig:15}. For DT and RF, the accuracy and F1-score of both models drop significantly on the testing dataset. As a result, SVM and ANN 
suffer less from overfitting while DT and RF suffer the most.

\begin{figure}[!htbp]
    \centering
    \subfloat[\centering Training time on CIC-IDS2017 dataset]{{\includegraphics[width=0.45\textwidth]{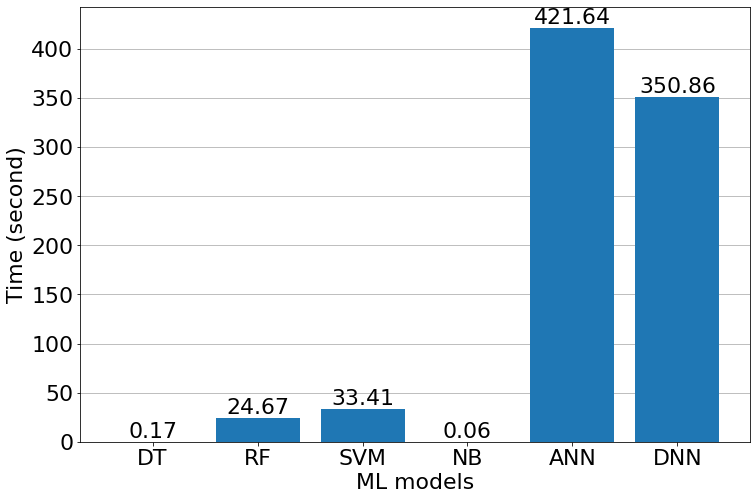} } \label{fig:16}}%
    \qquad
    \subfloat[\centering Prediction time on CSE-CIC-IDS2018 dataset]
    {{\includegraphics[width=0.45\textwidth]{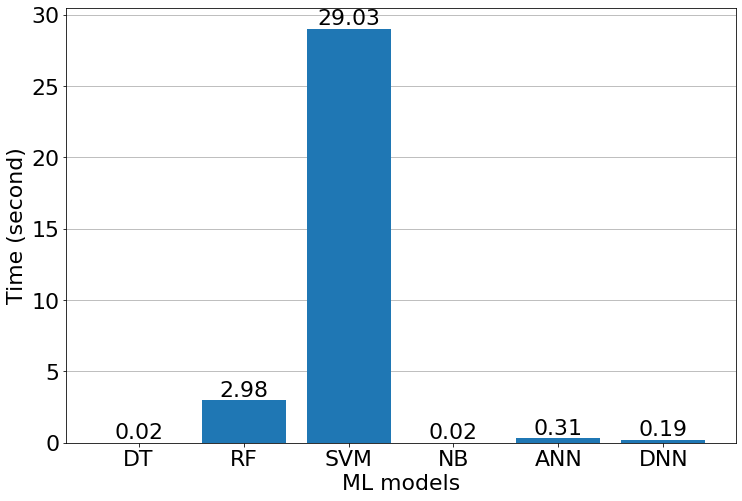} } \label{fig:17}}%
    \caption{Time consumption for training and prediction on the CIC dataset.}%
    \label{fig:16_&_17}%
\end{figure}

In terms of time consumption, ANN and DNN are costly to train. As shown in Figure~\ref{fig:16}, the training time of ANN and DNN is significantly longer than the other models implemented in this work. However, in terms of time consumption on classifying the samples, the time consumed by ANN and DNN are greatly reduced. As shown in Figure~\ref{fig:17}, SVM consumes significantly more time to predict the class of the samples. On the other hand, the time consumed by ANN and DNN to classify the samples is even less than that of the RF. DT and NB are the most efficient models in terms of time consumption for both training and predicting the class of the samples.
\subsection{Experiment Using LUFlow Dataset}
\label{sec:exp_luflow_dataset}
This section discusses the experiment details using the LUFlow dataset. For the experiment using the LUFlow dataset, the data collected in July 2020 is used as the training dataset, while the data collected in January 2021 is used as the testing dataset. For the rest of this paper, the dataset is referred by the year the data is collected. Hence, the training dataset is also referred to as LUFlow 2020 dataset, while the testing dataset is also referred to as LUFlow 2021 dataset. 
\subsubsection{Dataset Pre-Processing}
As the LUFlow dataset is organised according to the day that the data was collected, the first step is to combine the files of each day into one single file for processing. However, the datasets are huge in size; the LUFlow 2020 dataset contains 3.7 GB of data, while the LUFlow 2021 includes 2.5 GB of data. Hence, only 10\% of the samples are randomly selected for this work. 

We discovered that both datasets contain a small amount of missing value and duplicated samples. Besides that, no entry contain infinity values in both datasets. These unwanted entries are dropped as they only make up a small portion of the dataset. 

We also found that the datasets are slightly imbalanced. As shown in Table~\ref{tab:9}, the benign samples account for 56\% of the dataset. In contrast, the malicious traffic only accounts for 36\% of the dataset. Hence, benign samples are randomly selected so that the ratio of benign samples to malicious samples is 1:1. Moreover, samples in the class “outlier” are removed from the dataset as they are noise to the ML models. The class distribution of the datasets after cleaning and re-sampling is shown in Table~\ref{tab:10}.
\begin{table}[!htbp]
\centering
\caption{Class distribution of LUFlow dataset before cleaning.}
\label{tab:9}
\begin{tabular}{c|cc|cc}
\hline
\multirow{2}{*}{Classes} & \multicolumn{2}{c|}{LUFlow 2020} & \multicolumn{2}{c}{LUFlow 2021}\\ 
 & No. of rows (10\%) & No. of rows (\%) & No. of rows (10\%) & No. of rows (\%) \\ \thickhline
 benign	&1396168	&55.71\%	&1638952	&56.36\% \\ \hline
malicious	&905395	&36.12\%	&591372	&35.52\% \\ \hline
outlier	&204787	&8.17\%	&469345	&8.12\% \\ \hline
\end{tabular}
\end{table}
\begin{table}[!htbp]
\centering
\caption{Class distribution of LUFlow dataset after cleaning.}
\label{tab:10}
\begin{tabular}{c|cc|cc}
\hline
\multirow{2}{*}{Classes} & \multicolumn{2}{c|}{LUFlow 2020} & \multicolumn{2}{c}{LUFlow 2021}\\ 
 & No. of rows & No. of rows (\%) & No. of rows & No. of rows (\%) \\ \thickhline
 benign	&879740	&50\%	&569003	&50\% \\ \hline
malicious	&879740	&50\%	&569003	&50\% \\ \hline
%
\end{tabular}
\end{table}
\subsubsection{Feature Selection}
The LUFlow dataset is not large in terms of the number of features. The dataset only contains 16 features, where the description of each feature is listed in Table~\ref{tab:11}. Although there are only a few features, feature selection is still necessary as it would help remove noise from the dataset and improve the performance of the models. 
\begin{table}[!htbp]
\centering
\caption{Description of features of LUFlow dataset. Retrieved from reference \citep{mills2021practical}.}
\label{tab:11}
\begin{tabular}{c|l}
\hline
Feature & Description\\ \thickhline
src\_ip	& \makecell[l]{The source IP address associated with the flow. This feature is anonymised\\ to the corresponding Autonomous System.} \\ \hline
src\_port	& \makecell[l]{The source port number associated with the flow.} \\ \hline
dest\_ip	& \makecell[l]{The destination IP address associated with the flow. The feature is also \\anonymised in the same manner as before.} \\ \hline
dest\_port	& \makecell[l]{The destination port number associated with the flow.} \\ \hline
protocol	& \makecell[l]{The protocol number associated with the flow. For example, TCP is 6} \\ \hline
bytes\_in	& \makecell[l]{The number of bytes transmitted from source to destination.} \\ \hline
bytes\_out	& \makecell[l]{The number of bytes transmitted from destination to source.} \\ \hline
num\_pkts\_in	& \makecell[l]{The packet count from source to destination.} \\ \hline
num\_pkts\_out	& \makecell[l]{The packet count from destination to source.} \\ \hline
entropy	& \makecell[l]{The entropy in bits per byte of the data fields within the flow. This number \\ranges from 0 to 8.} \\ \hline
total\_entropy	& \makecell[l]{The total entropy in bytes over all of the bytes in the data fields of the flow.} \\ \hline
mean\_ipt	& \makecell[l]{The mean of the inter-packet arrival times of the flow.} \\ \hline
time\_start	& \makecell[l]{The start time of the flow in seconds since the epoch.} \\ \hline
time\_end	& \makecell[l]{The end time of the flow in seconds since the epoch.} \\ \hline
duration	& \makecell[l]{The flow duration time, with microsecond precision.} \\ \hline
label	& \makecell[l]{The label of the flow, as decided by Tangerine. Either benign, outlier, or \\malicious.} \\ \hline
\end{tabular}
\end{table}

We first removed some features manually before using the random forest to select the features. First, the ‘src\_ip’ and the ‘dest\_ip’ columns are removed as using the IP address to classify the traffics may cause overfitting in the long run. Besides that, the ‘time\_start’ and the ‘time\_end’ columns are also removed. The two columns record the start and end time using Unix timestamp, which will keep increasing over time. Besides that, the ‘duration’ column represents the time between ‘time\_start’ and ‘time\_end’. Hence, these two columns can be safely removed. 

\begin{figure}[!htbp]
\centering
\includegraphics[width=0.8\textwidth]{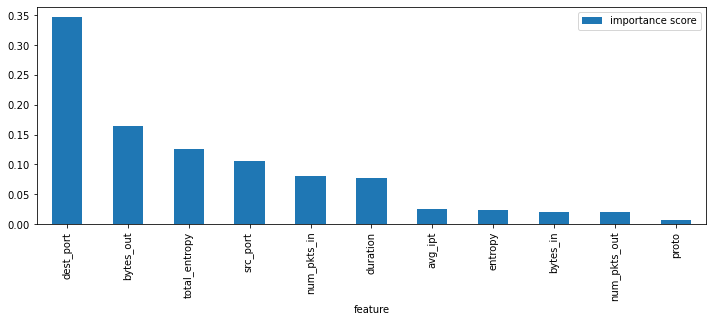}
\caption{Importance score of each feature in LUFlow dataset.}
\label{fig:18}
\end{figure}

After removing these four features, the remaining features are ranked according to their importance score computed by random forest. The ranking of the features is visualised using Figure~\ref{fig:18}. From Figure~\ref{fig:18}, ‘dest\_port’ is the highest scored, and its score is significantly higher than the ‘bytes\_out’ feature. Besides that, the importance scores of the last five features are substantially lower than the higher-ranked features. The low importance score indicates that those features do not provide much information to classify the data.

\begin{figure}[!htbp]
\centering
\includegraphics[width=0.9\textwidth]{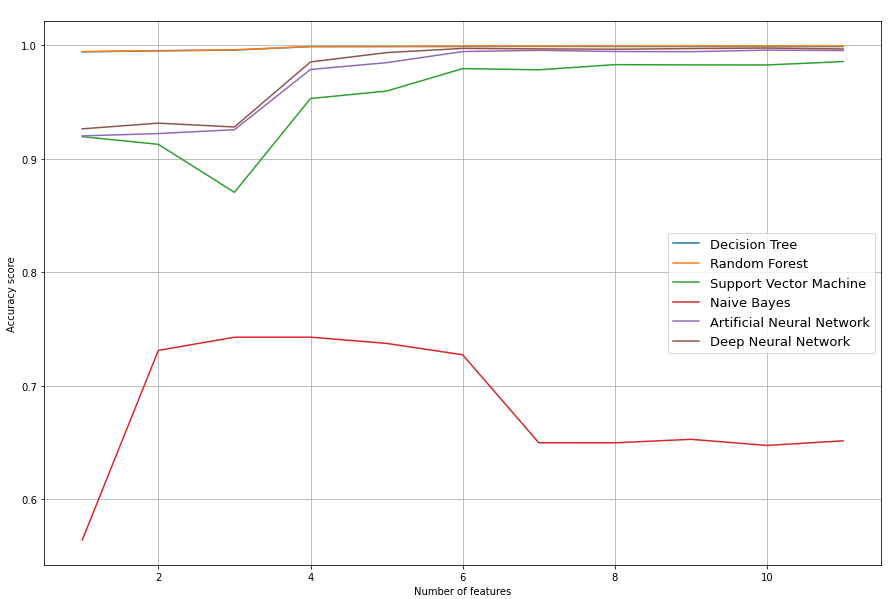}
\caption{Accuracy of the models with respect to the number of features in the LUFlow dataset.}
\label{fig:19}
\end{figure}

Finally, the features are further reduced using brute force. For implementation using the LUFlow dataset, all features are tested using brute force. Figure~\ref{fig:19} shows the accuracy of the models with respect to the number of features. The figure shows that naïve Bayes reach their maximum accuracy when the top two to six features are used. On the other hand, SVM, ANN and DNN achieved high accuracy when at least the top six features are included. For DT and RF, the accuracy is very high, even with only one feature. As a result, the top six features are chosen in the final feature set. 

\subsubsection{Accuracy Validation}
After optimising the hyperparameter of the models, it is important to check if the models are over-optimised, which will result in overfitting. We used five-fold cross-validation to validate the accuracies of the models. The accuracy of the models in each fold is listed in Table~\ref{tab:12}. We can see from Table~\ref{tab:12} that all models achieved excellent accuracies, except NB. Besides that, the standard deviation of the accuracy of each model is very minimal, indicating that the accuracies of the models are very consistent. 
\begin{table}[!htbp]
\centering
\caption{The accuracy result of 5-fold cross-validation on LUFlow dataset.}
\label{tab:12}
\begin{tabular}{c|c|c|c|c}
\hline
Model & Fold & Accuracy & Mean accuracy & Standard deviation \\ \thickhline
\multirow{5}{*}{Decision Tree}&
Fold-1 & 0.9995 & \multirow{5}{*}{0.9994} & \multirow{5}{*}{0.0001} \\
& Fold-2 &0.9993 &  & \\ 
& Fold-3 &0.9994 & & \\
& Fold-4 &0.9994 & & \\ 
& Fold-5 &0.9995 & & \\ \hline
\multirow{5}{*}{Random Forest}&
Fold-1 & 0.9995 & \multirow{5}{*}{0.9959} & \multirow{5}{*}{0.0001} \\
& Fold-2 &0.9994 &  & \\ 
& Fold-3 &0.9995 & & \\
& Fold-4 &0.9996 & & \\ 
& Fold-5 &0.9995 & & \\ \hline
\multirow{5}{*}{Support Vector Machine}&
Fold-1 & 0.9961 & \multirow{5}{*}{0.9959} & \multirow{5}{*}{0.0003} \\
& Fold-2 &0.9960 &  & \\ 
& Fold-3 &0.9962 & & \\
& Fold-4 &0.9956 & & \\ 
& Fold-5 &0.9954 & & \\ \hline
\multirow{5}{*}{Naïve Bayes}&
Fold-1 & 0.7201 & \multirow{5}{*}{0.7198} & \multirow{5}{*}{0.0026} \\
& Fold-2 &0.7154 &  & \\ 
& Fold-3 &0.7210 & & \\
& Fold-4 &0.7234 & & \\ 
& Fold-5 &0.7190 & & \\ \hline
\multirow{5}{*}{Artificial Neural Network}&
Fold-1 & 0.9953 & \multirow{5}{*}{0.9956} & \multirow{5}{*}{0.0003} \\
& Fold-2 &0.9960 &  & \\ 
& Fold-3 &0.9959 & & \\
& Fold-4 &0.9956 & & \\ 
& Fold-5 &0.9952 & & \\ \hline
\multirow{5}{*}{Deep Neural Network}&
Fold-1 & 0.9984 & \multirow{5}{*}{0.9984} & \multirow{5}{*}{0.0002} \\
& Fold-2 &0.9983 &  & \\ 
& Fold-3 &0.9983 & & \\
& Fold-4 &0.9987 & & \\ 
& Fold-5 &0.9986 & & \\ \hline
\end{tabular}
\end{table}

Looking at the accuracy score, DT and RF are the best models with an average of 99.94\% of accuracy. The worst performing model is NB, with only 72\% of accuracy, making it the least reliable model. In terms of accuracy, the performances of DT, RF and NB on the LUFlow dataset are aligned with those achieved using the CIC-IDS2017 dataset. On the other hand, the accuracies of SVM, ANN and DNN on the LUFlow dataset improve significantly, with accuracies above 99.5\%. 

For NB, it is the worst-performing model on both datasets. As described earlier, efforts have been made to improve the performance of NB. However, the accuracy of NB hardly improves any further. 
For the experiment using the LUFlow dataset, we do not compare the accuracy of each model with other literature. This is because the dataset is still very new, and there is no existing work using this dataset that can be compared to our work. Besides that, such comparison is unnecessary as most models have achieved higher than 99\% of accuracy. 
\subsubsection{Model Evaluation}
\begin{figure}[!tb]
\centering
\includegraphics[width=\textwidth]{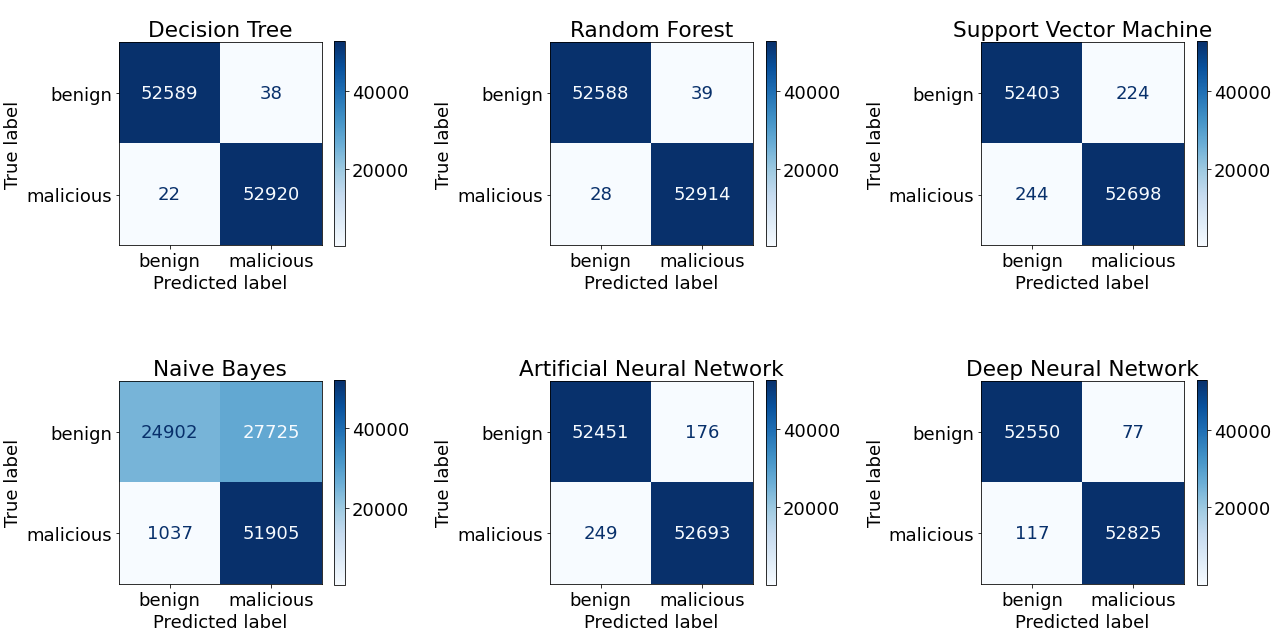}
\caption{Confusion matrix of each model on LUFlow 2020 dataset.}
\label{fig:20}
\end{figure}

Since we have ensured that the models are able to provide good accuracy by using cross-validation, we then test the model using the testing dataset, the LUFlow 2021 dataset. As described in Section~\ref{sec:model_evaluation}, the final ML models are trained using the LUFlow 2020 dataset with the optimal hyperparameters. %
As the size of the datasets is huge, only 20\% of the LUFlow 2020 and the LUFlow 2021 datasets are used in our experiments. We used 70\% of the LUFlow 2020 dataset for training the model; the other 30\% is used to evaluate the performance of the models on the training dataset. The performance of the models on the training dataset is listed in Table~\ref{tab:13} and visualised using confusion matrixes in Figure~\ref{fig:20}.
\begin{table}[!tb]
\centering
\caption{Performance of the models on LUFlow 2020 dataset.}
\label{tab:13}
\begin{tabular}{c|c|c|c|c|c}
\hline
\multirow{2}{*}{Models} & \multicolumn{4}{c|}{Evaluation Metrics} & \multirow{2}{*}{Class}\\ \cline{2-5}
& Accuracy & Precision & Recall & F1-score & \\ \thickhline
\multirow{2}{*}{Decision Tree} & \multirow{2}{*}{0.9994} & 0.9996 & 0.9993 & 0.9994 & benign \\
& & 0.9993 & 0.9996 & 0.9994 & malicious \\ \hline
\multirow{2}{*}{Random Forest} & \multirow{2}{*}{0.9994} & 0.9995 & 0.9993 & 0.9994 & benign \\
& & 0.9993 & 0.9995 & 0.9994 & malicious \\ \hline
\multirow{2}{*}{Support Vector Machine} & \multirow{2}{*}{0.9956} & 0.9954 & 0.9957 & 0.9956 & benign \\
& & 0.9958 & 0.9954 & 0.9956 & malicious \\ \hline
\multirow{2}{*}{Naïve Bayes} & \multirow{2}{*}{0.7276} & 0.9600 & 0.4732 & 0.6339 & benign \\
& & 0.6518 & 0.9804 & 0.7830 & malicious \\ \hline
\multirow{2}{*}{Artificial Neural Network} & \multirow{2}{*}{0.9960} & 0.9953 & 0.9967 & 0.9960 & benign \\
& & 0.9967 & 0.9953 & 0.9960 & malicious \\ \hline
\multirow{2}{*}{Deep Neural Network} & \multirow{2}{*}{0.9982} & 0.9978 & 0.9985 & 0.9982 & benign \\
& & 0.9985 & 0.9978 & 0.9982 & malicious \\ \hline
\end{tabular}
\end{table}

From Figure~\ref{fig:20}, we see that most models have achieved a good accuracy except NB. The problem of NB is that it overfits the benign samples, which is reflected in the recall score of NB on the malicious sample, with a score of only 0.4732. In other words, NB correctly classifies less than half of the benign samples. For other ML models, the precision and recall scores are very similar, indicating that they do not bias towards a specific class.

\begin{figure}[!tb]
\centering
\includegraphics[width=\textwidth]{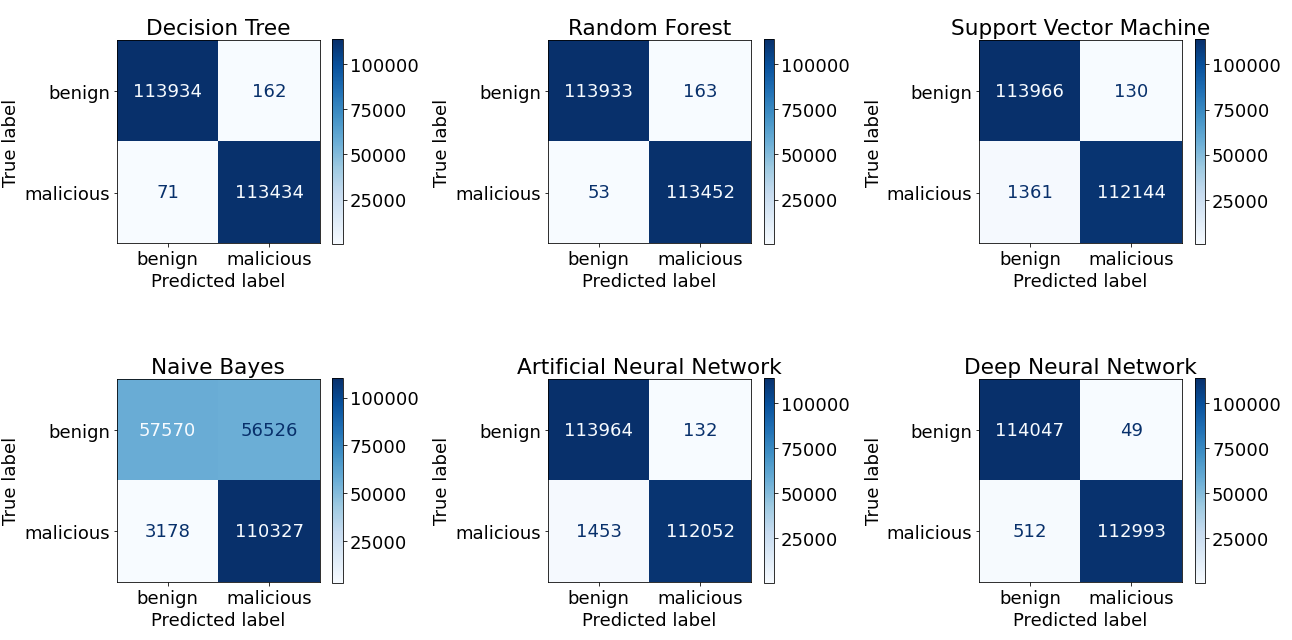}
\caption{Confusion matrix of each model on LUFlow 2021 dataset.}
\label{fig:21}
\end{figure}

The performance of the models on the LUFlow 2021 dataset is very interesting as the result is opposite to that of our first experiments on the CIC dataset. As shown in Figure~\ref{fig:21}, we notice that the models still perform very well on the LUFlow 2021 dataset. From the accuracy score of the models (see Table~\ref{tab:14}), we observe that the accuracy scores of most models do slightly decrease. For example, the accuracy score of the DT has fallen from 0.9994 to 0.9990, which is just a marginal drop. However, the recall score of NB on benign class has increased from 0.4732 to 0.5046, which is very surprising. 
\begin{table}[!tb]
\centering
\caption{Performance of the models on the LUFlow 2021 dataset.}
\label{tab:14}
\begin{tabular}{c|c|c|c|c|c}
\hline
\multirow{2}{*}{Models} & \multicolumn{4}{c|}{Evaluation Metrics} & \multirow{2}{*}{Class}\\ \cline{2-5}
& Accuracy & Precision & Recall & F1-score & \\ \thickhline
\multirow{2}{*}{Decision Tree} & \multirow{2}{*}{0.9990} & 0.9994 & 0.9986 & 0.9990 & benign \\
& & 0.9986 & 0.9994 & 0.9990 & malicious \\ \hline
\multirow{2}{*}{Random Forest} & \multirow{2}{*}{0.9991} & 0.9995 & 0.9986 & 0.9991 & benign \\
& & 0.9986 & 0.9995 & 0.9990 & malicious \\ \hline
\multirow{2}{*}{Support Vector Machine} & \multirow{2}{*}{0.9934} & 0.9882 & 0.9989 & 0.9935 & benign \\
& & 0.9988 & 0.9880 & 0.9934 & malicious \\ \hline
\multirow{2}{*}{Naïve Bayes} & \multirow{2}{*}{0.7377} & 0.9477 & 0.5046 & 0.6585 & benign \\
& & 0.6612 & 0.9720 & 0.7870 & malicious \\ \hline
\multirow{2}{*}{Artificial Neural Network} & \multirow{2}{*}{0.9930} & 0.9874 & 0.9988 & 0.9931 & benign \\
& & 0.9988 & 0.9872 & 0.9930 & malicious \\ \hline
\multirow{2}{*}{Deep Neural Network} & \multirow{2}{*}{0.9975} & 0.9955 & 0.9996 & 0.9975 & benign \\
& & 0.9996 & 0.9955 & 0.9975 & malicious \\ \hline
\end{tabular}
\end{table}

\begin{figure}[!tb]
    \centering
    \subfloat[\centering Accuracy of the models on LUFlow dataset]{{\includegraphics[width=0.7\textwidth]{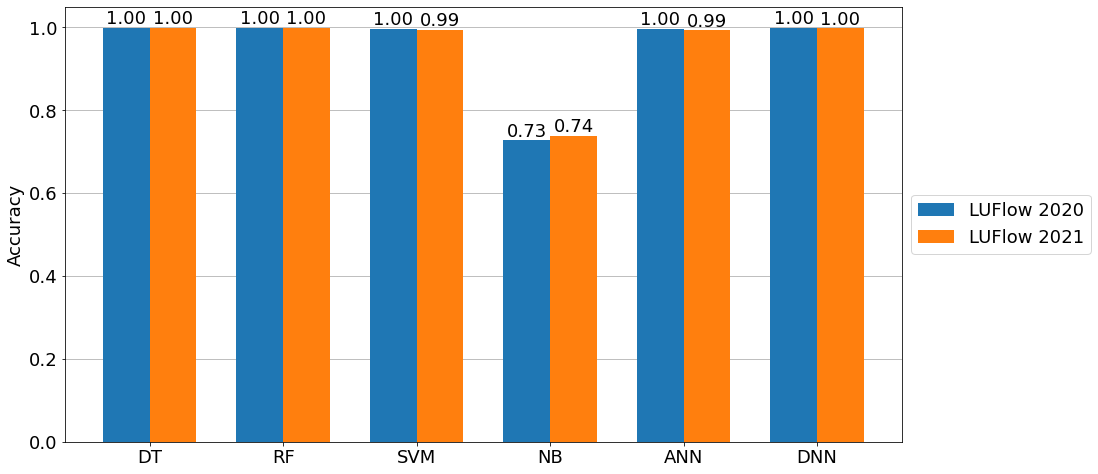} } \label{fig:22}}%
    \qquad
    \subfloat[\centering F1-score of the models on LUFlow dataset]
    {{\includegraphics[width=0.7\textwidth]{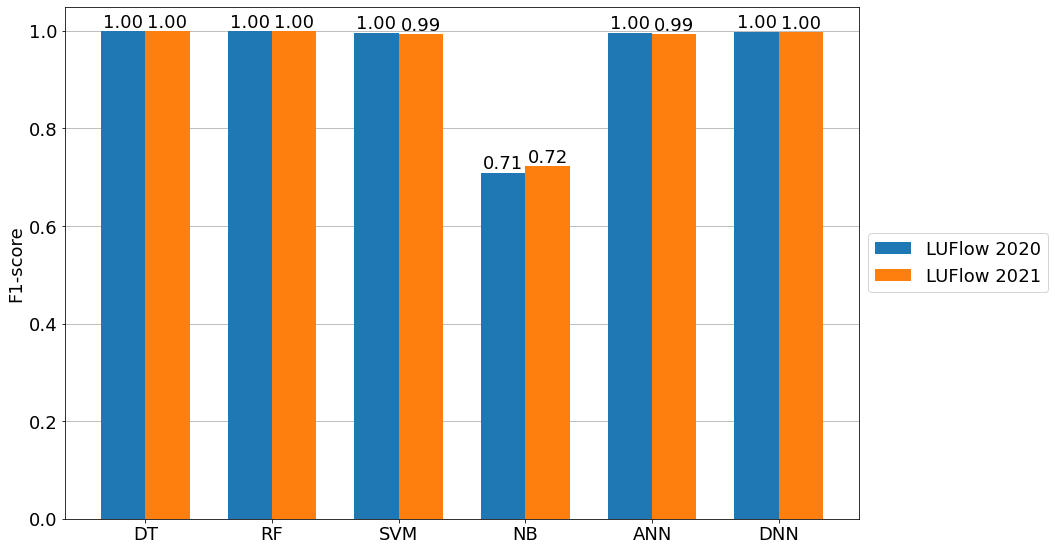} } \label{fig:23}}%
    \caption{Accuracy and F1-score of the models on the LUFlow dataset.}%
    \label{fig:22_&_23}%
\end{figure}

Figure~\ref{fig:22} and Figure~\ref{fig:23} provide further comparisons on the performance of the models on the training and testing dataset. From Figure~\ref{fig:22}, we observe that the accuracy of SVM and ANN dropped the most. Comparing the accuracy score shown in Table~\ref{tab:13} and Table~\ref{tab:14}, we see that the accuracy score of SVM and ANN have dropped more than 0.002. Besides that, the precision scores of both models on the benign class of the LUFlow 2021 dataset are lower than their recall score (the difference is about 0.01). The decrease in precision score may indicate that the models have started to bias towards the benign class. However, since the accuracy score and the F1-score of both models are still high, we cannot conclude that the models have started to bias towards a specific class. 

\begin{figure}[!tb]
    \centering
    \subfloat[\centering Training time on LUFlow dataset.]{{\includegraphics[width=0.45\textwidth]{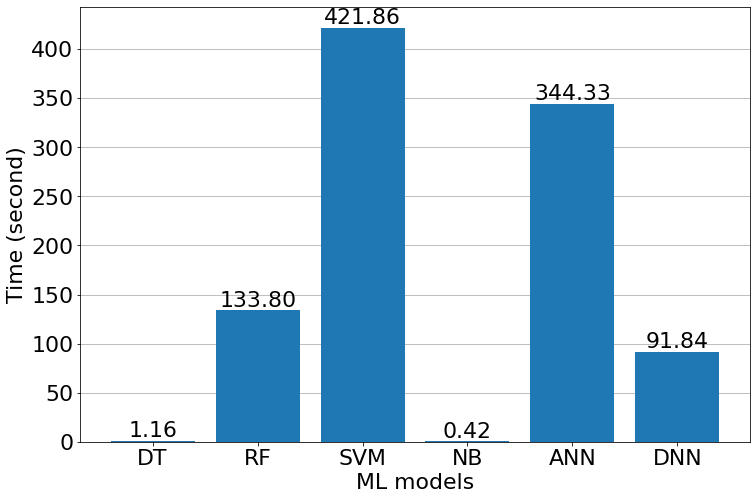} } \label{fig:24}}%
    \qquad
    \subfloat[\centering Prediction time on LUFlow dataset.]
    {{\includegraphics[width=0.45\textwidth]{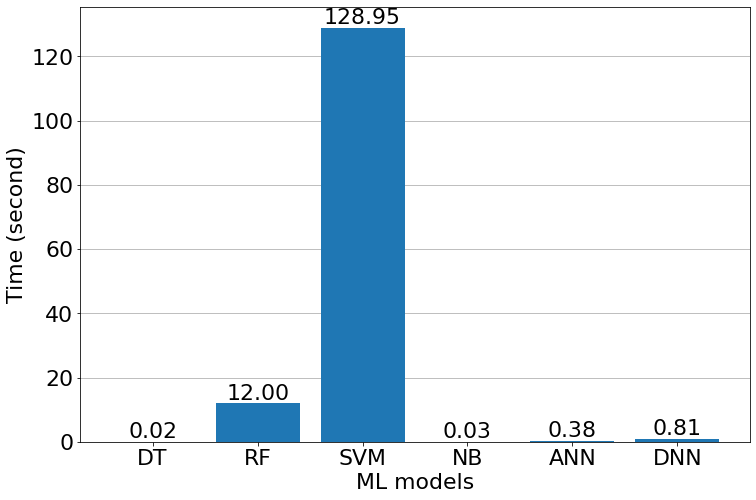} } \label{fig:25}}%
    \caption{Time consumption for training and prediction on LUFlow dataset.}%
    \label{fig:24_&_25}%
\end{figure}

In terms of time consumption for training, SVM and ANN are the most expensive models, as shown in Figure~\ref{fig:24}. 
On the other hand, DT and NB are still the most efficient models. Interestingly, DNN consumes less time to train than RF, SVM and ANN. Two reasons contribute to the reduction in training time. First, the number of neurons on each layer for the DNN is reduced to 10, compared to 15 neurons in the experiment using the CIC’s datasets. Besides that, the CIC’s dataset is wide, while the LUFlow dataset is long. In other words, the CIC’s datasets contain significantly more features than the LUFlow dataset, while the LUFlow dataset contains considerably more samples than the CIC’s dataset. Hence, the input layer contains fewer neurons. Besides that, the time complexity of each model is different. In terms of the time consumption for prediction, Figure~\ref{fig:25} shows that most models perform efficiently, while SVM consumes significantly more time than others.

\section{Discussion of Results}
\label{sec:discussion_of_results}
We conducted the experiment twice in this work, one using the CIC’s datasets and the other using the LUFlow dataset. Interestingly, the two experiments show two different results. The experiment results using the CIC’s datasets show that all models suffered from different degrees of overfitting. On the other hand, the experiment results using the LUFlow dataset show that the models do not have an overfitting problem. This difference comes from the fact that the two experiments reflect two different scenarios. 

The CIC’s datasets reflect a scenario where the environment used by the victims and the attackers have become more complex, and attackers are trying to break into the system using different tools and techniques. Hence, the deviation between the CIC-IDS2017 and the CSE-CIC-IDS2018 dataset is large. First, the CSE-CIC-IDS2018 dataset is created with more machines and various operating systems. Besides that, the class distribution of the CIC-IDS2017 and the CSE-CIC-IDS2018 dataset is different. For example, port scan attacks account for 14\% of the CIC-IDS2017 dataset, but there is no port scan attack in the CSE-CIC-IDS2018 dataset. On the other hand, bot attack and infiltration increase from 0.30\% and 0.01\% to 6\% and 3\%, respectively, in the CSE-CIC-IDS2018 dataset. The changes in the network environment and the type of attacks cause the models to perform poorly on the CSE-CIC-IDS2018 dataset. 

On the other hand, the LUFlow dataset reflects a scenario with minimum changes in the environment used by attackers and no changes in the environment used by the victims. Besides that, Lancaster University’s public address space may suffer fewer targeted attacks. Moreover, the type of attacks it faced may not change much within six months; we use data collected in July 2020 for the training dataset and use data collected in January 2021 for testing, which is six months apart. However, it is important to note that the LUFlow dataset contains 8\% of samples that are classified as an outlier. There may be some unseen attacks in the outlier class, but they are not correctly classified in the dataset.
 
The experiment results from the first set of experiments have shown that the method proposed in this paper has a better chance of detecting overfitting. From our experiment, we have shown that the models perform very well on the training dataset. We even evaluate the performance of the models on the training dataset using cross-validation, and there is no sign of overfitting. However, when we evaluate the models using the testing dataset, overfitting of the models is discovered. 

Our experiment results have also shown that ANN is the best model for a system for which the infrastructures are updated frequently, and the cost of cyberattack is very high. The ANN has balanced between resistance to overfitting and the time consumed on prediction. In contrast, DT is a better choice for a system that does not frequently update its infrastructure and does not suffer from massive attacks. This is because DT is one of the most efficient models in training and classifying new samples. Besides that, DT also provides the best accuracy on the training dataset in both experiments. It is also important to note that in the first set of experiments, the accuracy of the models on the testing dataset is lower than 80\%, which may be unacceptable for an IDS. Hence, periodically updating the IDS with newer data is necessary, regardless of the ML model being used.

\section{Conclusion}
\label{sec:conclusion}
We proposed a new method to evaluate the long-term performance of a machine learning (ML) based intrusion detection system (IDS).  Our proposed method uses a dataset created later than the training dataset to evaluate the ML models. To the best of our knowledge, this is the first work that trains and evaluates the ML-based IDS using separate datasets. We identified two sets of suitable datasets, the CIC dataset and the LUFlow dataset, to conduct our experiments using six ML models: decision tree (DT), random forest (RF), support vector machine (SVM), naïve Bayes (NB), artificial neural network (ANN) and deep neural network (DNN). 
From our experiment results, we conclude that ANN is the best model if long-term performance is important. On the other hand, DT is more suitable for small organisations that are less targeted to attacks. The experiment results also show that our proposed method can detect overfitting better. It is important to note that models proposed in other existing works may also suffer from the overfitting problem that occurred in our first experiment. However, the problem may not be discovered as the other existing works did not include a second dataset to evaluate their models. 

In future, we aim to evaluate unsupervised ML models and other more comprehensive models proposed by other literature. For example, we can evaluate one-class SVM using our proposed method as Hindy et al. \citep{hindy2020utilising} have shown that one-class SVM has good accuracy in detecting zero-day attacks. Besides that, future work may also aim to improve the feature selection method used in this paper. The selected features can greatly impact the performance of the models. Hence, it is possible that the overfitting observed in our first experiment can be improved by feature selection. 

\section*{Acknowledgements}
This work is supported by the Xiamen University Malaysia Research Fund under grants XMUMRF/2019-C3/IECE/0005 and XMUMRF/2022-C9/IECE/0032.

\bibliographystyle{elsarticle-num}

\end{document}

\end{document}